\documentclass[12, preprint]{aastex}   	% use "amsart" instead of "article" for AMSLaTeX format
\usepackage{geometry}                		% See geometry.pdf to learn the layout options. There are lots.
\usepackage{color}
\usepackage{natbib}
\usepackage{graphicx}				% Use pdf, png, jpg, o with pdflatex; use eps in DVI mode
\usepackage{amssymb}
\usepackage{calc}
\usepackage{lscape}
\usepackage{multirow}

\begin{document}

\title{The Vertical Metallicity Gradient of the Milky Way Disk: Transitions in [$\alpha$/Fe] Populations}

\author{Katharine J.~Schlesinger\altaffilmark{1},
Jennifer A.~Johnson\altaffilmark{2}, 
Constance M.~Rockosi\altaffilmark{3},
Young Sun Lee\altaffilmark{4},
Timothy C. Beers\altaffilmark{5},
Paul Harding\altaffilmark{6},
Carlos Allende Prieto\altaffilmark{7,8},
Jonathan C.~Bird\altaffilmark{9}
Ralph Sch{\"o}nrich\altaffilmark{10},
Brian Yanny\altaffilmark{11},
Donald P.~Schneider\altaffilmark{12,13},
Benjamin A.~Weaver\altaffilmark{14},
Jon Brinkmann\altaffilmark{15} 
}

\altaffiltext{1}{Research School of Astronomy and Astrophysics, Mount Stromlo Observatory, The Australian National University, Acton, ACT 2611 Australia} 
\altaffiltext{2}{Department of Astronomy, The Ohio State University, 140 W 18th Ave, Columbus, OH 43210 USA}
\altaffiltext{3}{UCO/Lick Observatory, University of California, Santa Cruz, CA 95064 USA}
\altaffiltext{4}{Department of Astronomy, New Mexico State University, Las Cruces, New Mexico 88003 USA; Tombaugh Scholar}
\altaffiltext{5}{National Optical Astronomy Observatory, Tucson, AZ, 85737 USA and JINA: Joint Institute for Nuclear Astrophysics}
\altaffiltext{6}{Department of Astronomy, Case Western Reserve University, Cleveland, OH 44106 USA}
\altaffiltext{7}{Instituto de Astrof\'{i}sica de Canarias, 38205 La Laguna, Tenerife, Spain}
\altaffiltext{8}{Departamento de Astrof\'{i}sica, Universidad de La Laguna, 38206 La Laguna, Tenerife, Spain} 
\altaffiltext{9}{Department of Physics and Astronomy, Vanderbilt University, 6301 Stevenson Center, Nashville, TN 37235 USA ; VIDA Postdoctoral Fellow}
\altaffiltext{10}{Rudolf-Peierls Centre for Theoretical Physics, University of Oxford, 1 Keble Road, OX1 3NP, Oxford, United Kingdom}
\altaffiltext{11}{Fermi National Accelerator Laboratory, P.O. Box 500, Batavia, IL 60510 USA}
\altaffiltext{12}{Department of Astronomy and Astrophysics, The Pennsylvania State University, University Park, PA 16802 USA} 
\altaffiltext{13}{Institute for Gravitation and the Cosmos, The Pennsylvania State University, University Park, PA 16802 USA}
\altaffiltext{14}{Center for Cosmology and Particle Physics, New York University, New York, NY 10003 USA}
\altaffiltext{15}{Apache Point Observatory, Sunspot, NM 88349 USA} 

\begin{abstract} 
Using G dwarfs from the Sloan Extension for Galactic Understanding and Exploration (SEGUE) survey, we have determined a vertical metallicity gradient over a large volume of the Milky Way's disk, and examined how this gradient varies for different [$\alpha$/Fe] subsamples. This sample contains over 40,000 stars with low-resolution spectroscopy over 144 lines of sight. We employ the SEGUE Stellar Parameter Pipeline (SSPP) to obtain estimates of effective temperature, surface gravity, [Fe/H], and [$\alpha$/Fe] for each star and extract multiple volume-complete subsamples of approximately 1000 stars each. Based on the survey's consistent target-selection algorithm, we adjust each subsample to determine an unbiased picture of the disk in [Fe/H] and [$\alpha$/Fe]; consequently, each individual star represents the properties of many. The SEGUE sample allows us to constrain the vertical metallicity gradient for a large number of stars over a significant volume of the disk, between $\sim$0.3 and 1.6 kpc from the Galactic plane, and examine the \emph{in situ} structure, in contrast to previous analyses which are more limited in scope. This work does not pre-suppose a disk structure, whether composed of a single complex population or a distinct thin and thick disk component. The metallicity gradient is $-$0.243$^{+0.039}_{-0.053}$ dex kpc$^{-1}$ for the sample as a whole, which we compare to various literature results. Each [$\alpha$/Fe] subsample dominates at a different range of heights above the plane of the Galaxy, which is exhibited in the gradient found in the sample as a whole. Stars over a limited range in [$\alpha$/Fe] show little change in median [Fe/H] with height. If we associate [$\alpha$/Fe] with age, our consistent vertical metallicity gradients with [$\alpha$/Fe] suggest that stars formed in different epochs exhibit comparable vertical structure, implying similar star-formation processes and evolution. 
\end{abstract}

\section{Introduction}
\label{sec:intro} 

Understanding the structure and evolution of the Milky Way disk is imperative for developing a complete picture of galaxy formation in a $\Lambda$CDM universe. The variation in chemistry throughout the disk provides information about the star-formation history and stellar dynamics and how they change as the Galaxy evolves. As we can resolve individual stars in the Milky Way, it provides the best opportunity to examine the chemical structure of the Galactic disk in detail. Previous analyses of external galaxies \citep{burstein79, dalcanton02, yoachim08a, yoachim08b}, including those at redshifts as high as $z\sim$3 \citep{elmegreen06}, find that our Galaxy's global structure is commonplace, making the Milky Way a Rosetta Stone for galaxy-formation processes. 

Previous observations of the Milky Way disk were interpreted in the context of two distinct components, a thin and thick disk. The ``thick disk" was first identified by \citet{gilmore83}, who found that the stellar number density as a function of height above the plane was best fit by two components\footnote[1]{The likely existence of a second stellar component close to the Galactic plane was previously reported by \citet{yoshii82}, who referred to it as a halo component, even though its inferred density was 10 times that of the local halo.  The \citet{yoshii82} normalization relative to the local thin disk (0.01--0.02) and scale height ($\sim$2 kpc) were commensurate with the values later determined by \citet{gilmore83} (0.3, and 1.35 kpc, respectively). See also \citet{yoshii87} and \citet{yoshii13}.}, one with a scale height of approximately 300 pc, and the second with a scale height of 1350 pc. Work by \citet{juric08} later revised these values to 300 and 900 pc respectively. 

When a star forms, it is imprinted with information about the chemistry of its natal cloud. Specifically, chemistry reveals the properties of the dying stars that enriched the surrounding material and the rate of star formation. For example, older stars are largely enriched by core-collapse supernovae (SNe II), which occur on timescales of 10$^{7}$ years and release $\alpha$-elements (e.g., O, Mg, Si, Ca, Ti) to their surroundings (e.g., \citealt{arnett78}, \citealt{woosley95}). These stars are metal-poor with relatively high values of [$\alpha$/Fe]. After $\approx$10$^8$ years, Type Ia supernovae (SNe Ia) occur (e.g., \citealt{smeckerhane92}), releasing Fe-peak elements into the surrounding material (e.g., \citealt{nomoto84}). Younger stars will thus have lower [$\alpha$/Fe] and higher [Fe/H]. Although the dynamics of the star may change over its lifetime, the stellar chemistry remains largely the same \citep{freeman}. We can thus use stars as a fossil record of earlier Galaxy conditions. 

In addition to star counts, the thin and thick disks are apparent as bimodal distributions in stellar kinematics and chemical composition. When separated by kinematics (e.g., \citealt{soubiran03}), the thin disk is metal-rich (mean [Fe/H]$\approx-$0.2) and $\alpha$-poor, while the thick disk is relatively metal-poor (mean [Fe/H]$\approx-$0.6) and enhanced in [$\alpha$/Fe] \citep{gilmore85, wg95, chiba00, fuhrmann98, fuhrmann11, prochaska00, bensby03, bensby05, reddy06}. This chemical variation is consistent with observations that thick-disk stars are generally older than thin-disk stars, with ages of around 8 Gyr and higher \citep{bensby03, bensby05, reddy06, haywood13}. Instead of kinematically defining the thin and thick disk, \citet{lee11b} first grouped disk stars in chemical space, finding that the $\alpha$-rich and $\alpha$-poor subsamples exhibited different gradients of rotational velocity with metallicity, supporting a picture of a Galactic disk with two separable components. 

The varying chemistry, in conjunction with differing kinematics and age, suggest that the thin and thick disk had distinct star-formation histories and evolution. However, other studies do not support this two-component paradigm. \citet{norris_ryan} questioned whether or not the two components were actually separable from one another. Analysis of the photometric Sloan Digital Sky Sample (SDSS; \citealt{york00}) sample by \citet{ivezic08}, supported this idea, finding that the metallicity distribution of the disk was best modelled by a complex single  component. In addition, \citet{schonrich09a} suggest that the bimodality in [$\alpha$/Fe] can occur in a single stellar population because the [$\alpha$/Fe]-enhancement fades rapidly once SNe Ia enrichment becomes important; this separation also occurs naturally in the chemodynamical models of \citet{minchev13}. Recent work by \citet{bovy12a, bovy12b} determined that disk stars exhibit smooth trends between their chemical composition ([$\alpha$/Fe] and [Fe/H]) and kinematics, proposing that the Milky Way disk represents a continuum of stellar populations, rather than two separable structures. 

There is clear variation chemically and dynamically over the disk, which suggests multiple populations. However, it is currently unclear the degree to which the criteria used to divide the disk define our picture of it. Specifically, are there distinct thin- and thick-disk components? Or have our adopted analysis techniques artificially divided a single complex population by presupposing a two-component solution? By examining how the stellar chemistry varies over a large volume of the disk, without assuming any particular disk structure, we can investigate the star-formation history of the disk as a whole and search for thin/thick disk distinctions. 

Any promising model of the dynamical formation and chemical evolution of the Milky Way's disk must recreate the observed physical and chemical structure, specifically how the chemistry varies with respect to location. In particular, the variation in metallicity with respect to distance from the Galactic plane is a valuable constraint for different models of disk development, in addition to providing insight into the proposed thin/thick disk dichotomy. Furthermore, as these metallicity gradients can be measured in both external galaxies (e.g., \citealt{degrijs}) and simulations (e.g., \citealt{loebman11}), they can place the Milky Way in the larger context of Galactic evolution.

Thick-disk formation via stellar accretion, as explored by \citet{abadi03}, produces no metallicity gradient with respect to distance from the plane, as the final $|Z|$ value of the stars does not depend on their [Fe/H]. \citet{brook04, brook05} and \citet{bournaud09} propose that the thick disk may result from early accretion of gas-rich material. This accretion prompts a burst in star formation, resulting in a well-mixed, chemically uniform thick disk. Finally, work by \citet{roskar08}, \citet{schonrich09a}, and \citet{loebman11} propose that the thick disk forms via radial migration, whether by processes internal to the galaxy or enhanced by mergers or close encounters \citep{quillen09, bird12}. This process involves stars moving radially from the inner to the outer regions of the disk, and vice versa, due to scattering by transient spiral structure \citep{sellwood02} or diffusion caused by overlap of bar-spiral resonances \citep{minchev10, brunetti11, minchev11}. As stars move outward, they experience a lower gravitational restoring force in the outer disk, allowing them to move away from the midplane. However, models of how radial migration affects Galactic structure vary significantly (e.g., \citealt{schonrich09a}, \citealt{loebman11}, \citealt{minchev12}); in particular, assumed parameters, such as the star formation history, can drastically affect observable predictions, like the vertical metallicity gradient.

Questions of the structure and development of the Milky Way disk emphasise the need for an unbiased stellar sample that extends over a large volume of the disk, such that we can investigate the metallicity structure over both proposed disk components with a uniform data set. The long lifetimes of cool dwarf stars make them a valuable fossil record of Galactic evolution; their atmospheric chemical composition is largely unchanged since birth and reflects the chemical makeup of the gas from which the star formed. Some of them are surviving relics from the earliest epochs of the Milky Way. However, their low luminosity makes it difficult for spectroscopic analyses to extend far beyond the solar neighborhood. Previous analyses of the vertical metallicity structure of the Galactic disk have had to compromise between sample size, volume coverage, and accuracy of stellar parameters and distances. 

\citet{hartkopf82} examined a sample of around 1000 G and K dwarfs towards the North and South Galactic pole with David Dunlap Observatory photometry and reported a vertical metallicity gradient of $-$0.2 dex kpc$^{-1}$ for stars with $|Z|<$5 kpc. Similarly, \citet{ak07} measured a vertical metallicity gradient of $-$0.20$\pm$0.02 dex kpc$^{-1}$ for stars with $|Z|<$3 kpc over two lines of sight at $l$ of 180$^{\circ}$ and $b$ of $\pm$45$^{\circ}$ using photometric metallicities. \citet{ivezic08} greatly expanded the scope of these analyses, finding similar behavior using a photometric sample of over two million stars covering an unprecedented volume of the Galaxy. Unfortunately, these analyses are limited by their reliance on photometric metallicity indicators, which are susceptible to errors from reddening corrections, have reduced sensitivity at low metallicity, and depend strongly on the adopted calibration to spectroscopic estimates, which vary from work to work. Furthermore, these analyses typically rely on photometric parallaxes for stellar distances, which can have large uncertainties.

Spectroscopic analyses (e.g., \citealt{katz11}; \citealt{kordopatis11}) can determine more accurate chemical abundances, and provide radial velocity information, but they require increased observing time. This restricts most spectroscopic samples to hundreds of stars over a small number of lines of sight. In addition, high-resolution studies of disk stars (e.g., \citealt{adibekyan}; \citealt{bensby03}; \citealt{bensby05}) are not able to probe the \emph{in situ} thick disk, making it difficult to adequately constrain the vertical metallicity gradient over a significant volume. Differences in spatial coverage, sample selection, and analysis techniques for the spectroscopic analyses manifest themselves as significant variation in their vertical metallicity gradient estimates, with deviations larger than the reported uncertainties. Many spectroscopic studies measure gradients that differ from both photometric studies and each other. For example, while \citet{katz11} and \citet{allendeprieto06} measure a small thick-disk vertical metallicity gradient of $-$0.068 dex kpc$^{-1}$ and 0 dex kpc$^{-1}$ respectively, \citet{kordopatis11} estimate a gradient of $-$0.14 dex kpc$^{-1}$.  Furthermore, many of these analyses distinguish the thin and thick disk components, either kinematically or geometrically, rather than examining the disk as a whole, making it difficult to determine if the distinctions between components are ``real" or a consequence of the assumed separation. 

Numerous surveys have sought to address the need for a significant sample over a large volume with accurate stellar parameters. The Geneva-Copenhagen Survey has tens of thousands of stars with well-constrained photometric metallicities (e.g., \citealt{casagrande11}). However, it is only volume-complete within 40 pc, and thus, does not extend far beyond the solar neighborhood. The more recent RAdial Velocity Experiment (RAVE) has around 17,000 F and G dwarfs and spectroscopic metallicities \citep{siebert11}. The dwarfs in the RAVE sample probe to a maximum distance of approximately 1 kpc; the typical distance for their cool dwarf sample ranges from 50--250 pc \citep{zwitter10, steinmetz12}. Thus, their dwarf sample is not ideal for determining the \emph{in situ} vertical metallicity gradient, as it does not extend far beyond the plane of the Galaxy, which is dominated by metal-rich stars. Recent work by \citet{hayden13} on the first year of data from the Sloan Digital Sky Survey (SDSS) III Apache Point Observatory Galactic Evolution Experiment (APOGEE; \citealt{ahn13}) experiment is quite promising; they currently have observed more than 80,000 stars over a large range of Galactocentric radius and height and have publicly released data for around 30,000 targets. As this survey continues, it will provide valuable insight into the Galactic structure, with observations of over 100,000 stars. Similarly, the Galactic Archaeology with HERMES (GALAH; \citealt{zucker12}) survey will provide high-resolution spectra for over a million stars, enabling examination of the variation over the disk for up to 25 different elements over a magnitude-limited volume. 

In this work, we use the G-dwarf stars from the Sloan Extension for Galactic Understanding and Exploration (SEGUE; \citealt{yanny09}) survey to determine the vertical metallicity gradient of the disk over a wide range of heights above the Galactic plane, from around 0.3 to 1.6 kpc. We also compare our vertical metallicity gradients directly to simulations and other observational studies, a useful comparison with and without [$\alpha$/Fe] information. SEGUE provides SDSS $ugriz$ photometry \citep{fuku96} and low-resolution spectroscopy for 240,000 stars over a range of 14$<g<$20.3 in $\sim$3500 square degrees on the sky. Not only is this the largest spectroscopic sample currently available, it covers a much more extensive volume of the Milky Way disk than all previous analyses. SEGUE utilises consistent and well-documented selection criteria, allowing for correction of the different observational biases in the sample, such that it accurately reflects the underlying disk structure. The survey also provides [Fe/H] and [$\alpha$/Fe] information for each individual star, enabling the determination of an unbiased vertical metallicity gradient of the disk beyond a local volume and examination of its variation with respect to $\alpha$-element abundance ratios. 

The paper is organised as follows. We first discuss assembly of our program sample, and our corrections for observational biases, in \S\,\ref{sec: sample}. We also consider different ways of dividing the sample, in order to investigate the disk chemical structure for different subpopulations. \S\,\ref{sec: vmg} presents our technique for determining the vertical metallicity gradient and the various uncertainties that factor into our calculations. We present the measured gradients for our various subsamples in \S\,\ref{sec: vmg_results}. In \S\,\ref{sec: discussion}, we compare our values to the results from previous work and examine our gradients in the context of different Galaxy formation models and their predictions for the vertical chemical structure. Finally, we summarise our results in \S\,\ref{sec:summary}.
      
\section{The SEGUE G-dwarf Sample} 
\label{sec: sample}

\begin{figure}[htbp]
\begin{center}
\includegraphics[width=\textwidth]{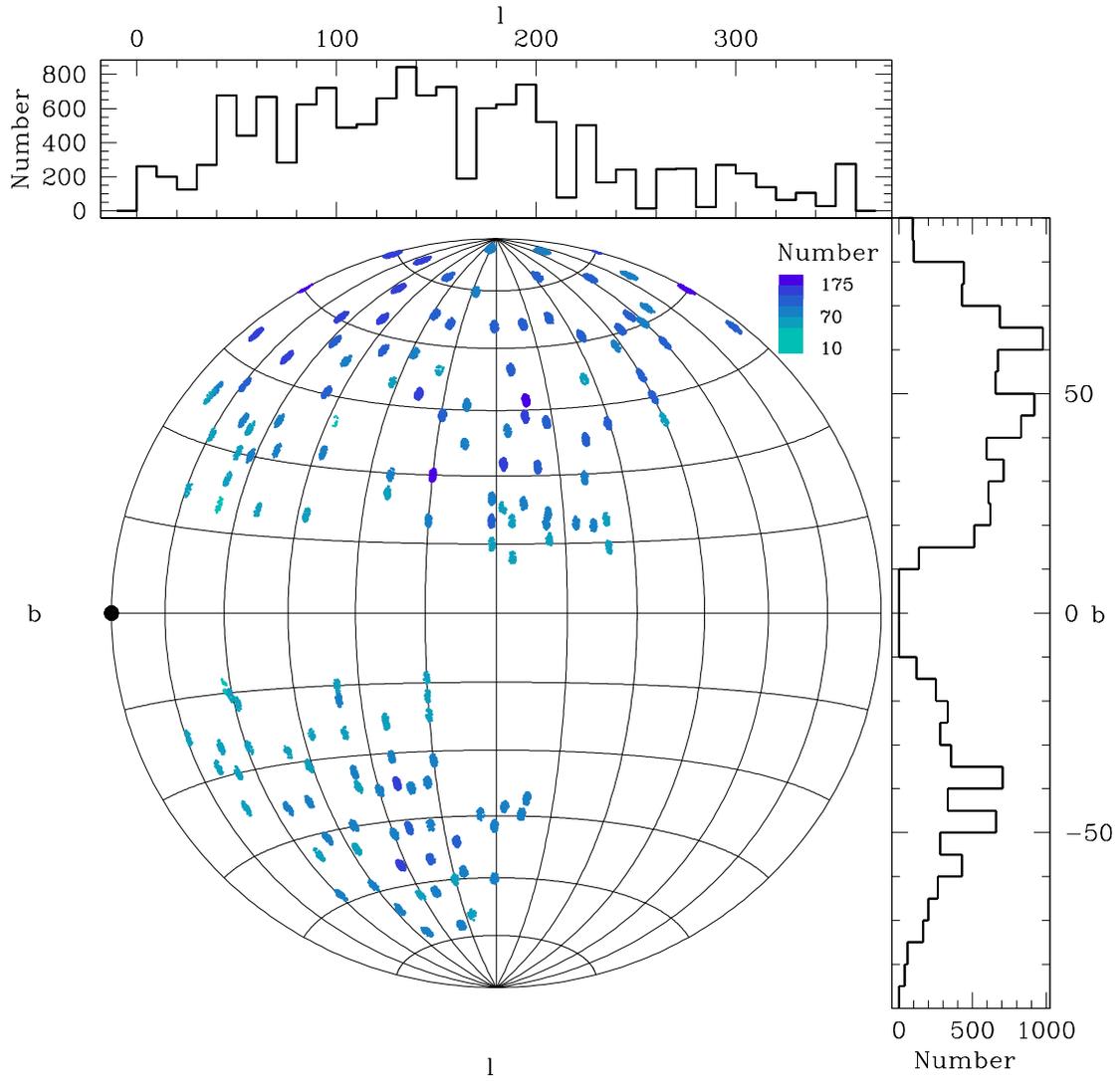}
\caption{Coverage of the SEGUE G-dwarf sample in Galactic coordinates. The circular black point indicates the position of the Galactic Center. The top histogram shows the distribution of the sample in Galactic longitude, $l$; the right hand side shows the distribution in Galactic latitude, $b$. The different colors represent the number of G stars in our sample from that region.
}
\label{fig:spatial_alpha_aitoff}
\end{center}
\end{figure}

The SEGUE G-dwarf sample is defined by simple color (0.48$<(g-r)_0<$0.55) and magnitude (14.0$<r_0<$20.2) criteria \citep{yanny09}\footnote[2]{These color cuts are equivalent to a spectral type of G5 or G6 \citep{johnson63}.}. The subscript 0 indicates dereddening and extinction correction using \citet{sfd98} values. As shown in Figure\,\ref{fig:spatial_alpha_aitoff}, this sample consists of 144 lines of sight with Galactic latitudes $b > 10^{\circ}$. An in-depth discussion of the SEGUE survey and target-selection design are provided in \citet{schlesinger12}.  Technical information about SDSS and SEGUE is published on the survey design \citep{york00, eisenstein11}, telescope and camera \citep{gunn06, gunn98}, astrometric \citep{pier03} and photometric \citep{ivezic04} accuracy, photometric system \citep{fuku96}, photometric calibration \citep{hogg01, smith02, tucker06, pad08}, and recent updates to the instrumentation \citep{smee13}. 

Since the publication of \citet{schlesinger12}, SDSS Data Release 9 (DR9; \citealt{ahn12}) has provided improved photometry and an updated version of the stellar parameters determined by the SSPP. We extract an updated sample of 42,901 stars with SEGUE spectroscopy that meet the G-dwarf selection criteria in SDSS DR9 photometry. Rather than containing new SEGUE data, DR9 provides a reprocessing of the SEGUE observations. 

To avoid saturation effects, we trim our sample to only include stars with $r > $15. We also remove lines of sight with extinction greater than 0.5 mags in \emph{r} and E$(g-r)$ of greater than 0.2 mag. These criteria, and other concerns of the effect of extinction on the G-dwarf sample, are discussed at length in \citet{schlesinger12}. To ensure spectroscopic quality, we avoid stars that are flagged due to temperature or noise issues. We also require that all stellar spectra have an average S/N$ > $30 per pixel\footnote[3]{Each stellar target is observed with spectral resolution of 1800 from 3800 to 9200\AA\, \citep{eisenstein11}.}; at this quality the SSPP uncertainties in [Fe/H] and [$\alpha$/Fe] are 0.23 and 0.1 dex, respectively. 

We use SSPP $\log g$ values to ensure that our targets are dwarfs. \citet{schlesinger12} used an optimised version of the SSPP DR8 pipeline, removing individual $\log g$ determinations that were not well-suited for cool dwarfs. The SSPP for DR9 takes into account this analysis of the DR8 pipeline, removing the individual techniques that are inaccurate or inappropriate. Thus, we no longer need to use a modified version of the SSPP stellar parameters, and our sample is replicable using the SDSS data-mining infrastructure.

\begin{figure}[htbp]
\begin{center}
\includegraphics[width=\textwidth]{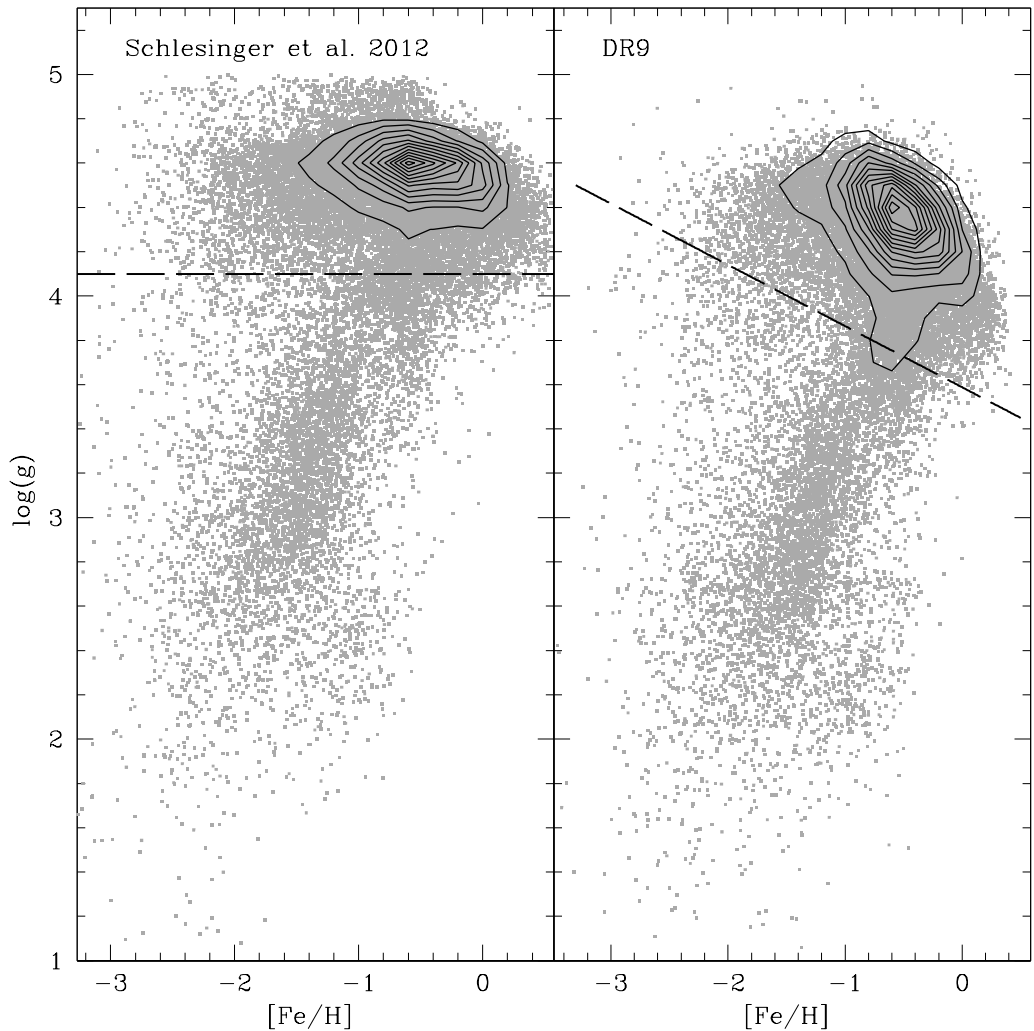}
\caption{Modifications to the SSPP for DR9 have changed the distribution of $\log g$ with respect to [Fe/H] for SEGUE stars. The grey points are individual stars that meet the G-dwarf color and magnitude criteria. The black contours reflect the density of stars, in increments of 250. The left panel shows the distribution for the parameters from the optimised DR8 SSPP G-dwarf sample used by \citet{schlesinger12}. A straight cut at $\log g =$ 4.1, the dashed black line, was used to separate dwarf and giant stars. The right panel displays the atmospheric parameters in the DR9 SSPP. The distribution is now angled with respect to [Fe/H]; we must use a sloping line to isolate dwarf stars, as shown by the dashed black line. 
}
\label{fig:logg_distribution}
\end{center}
\end{figure}

There were also changes made to the SSPP to improve atmospheric parameter estimates for more metal-rich and hot stars; these adjustments affected the estimates for the G-dwarf surface gravities. The sample now exhibits a systematic decrease in $\log g$ with increasing [Fe/H] (Figure\,\ref{fig:logg_distribution}). We use the SSPP $\log g$ values only as a diagnostic to ensure our sample consists of dwarf stars, thus we are not concerned about the systematic behavior. 

Previously, we removed evolved stars with a cut at $\log g$ of 4.1. Using the Mg index, \citet{schlesinger12} determined that this limit in $\log g$ resulted in a K-dwarf sample with less than 1\% contamination from evolved stars. Due to the target-selection design of SEGUE, we expect that contamination by evolved stars in the G-dwarf sample would be even smaller. For DR9, we must use a sloping $\log g$ cut to isolate dwarf stars (Figure\,\ref{fig:logg_distribution}). We define a line from $\log g$ of 4.5 at [Fe/H] $= -3.3$ to $\log g$ of 3.45 at [Fe/H] $= +0.5$.  Stars that fall below this line are removed from the sample. Fewer than 2\% of stars that met the original $\log g$ criteria in \citet{schlesinger12} are removed from the sample using our updated DR9 cuts. Our new $\log g$ criteria reflects a shift in the distribution; our sample of stars remains largely the same. Thus, as \citet{schlesinger12} found negligible contamination of the G-dwarf sample by giants and subgiants, there is little contamination by evolved stars of our updated sample. 

\subsection{Removing Halo Stars} 

Our SEGUE sample has a number of metal-poor stars that may be interlopers from the halo system. These stars can artificially shift the median [Fe/H] to lower values, affecting our measured vertical metallicity gradient. Limiting our sample to stars with [Fe/H]$>-$1.0 avoids contamination from halo stars, but it also may remove evidence of a metal-weak thick disk from our sample. As these stars are a small portion of our sample, their removal does not significantly change the determined vertical metallicity gradient. Thus, we limit our sample to stars with [Fe/H] above $-$1.0. 

\subsection{Distance Determinations}

For each star that passes SEGUE target selection and our spectral-quality criteria, we determine the distances by matching them in $(g-r)_{0}$ and [Fe/H] to 10 Gyr isochrones from the empirically-corrected Yale Rotation Evolution Code (YREC) set \citep{an09}. This technique, and an in-depth analysis of associated uncertainties, are described in detail in Schlesinger et al. (2012). Briefly, there are random distance errors arising from uncertainties in photometry, SSPP estimates of [Fe/H] and [$\alpha$/Fe], and isochrone choice. The total random distance error is dominated by uncertainties in SSPP [Fe/H] estimates and ranges from around 18$\%$ for stars with [Fe/H] $> -$0.5 to 8$\%$ for more metal-poor stars. There are also systematic distance uncertainties from using 10 Gyr isochrones and the possibility of undetected binarity; these two uncertainties have a negligible effect on the distance of metal-poor stars, while the most metal rich stars have a systematic shift in distance of $-$3$\%$, which is factored into our distance estimates. 

\begin{figure}[htbp]
\begin{center}
\includegraphics[width=\textwidth]{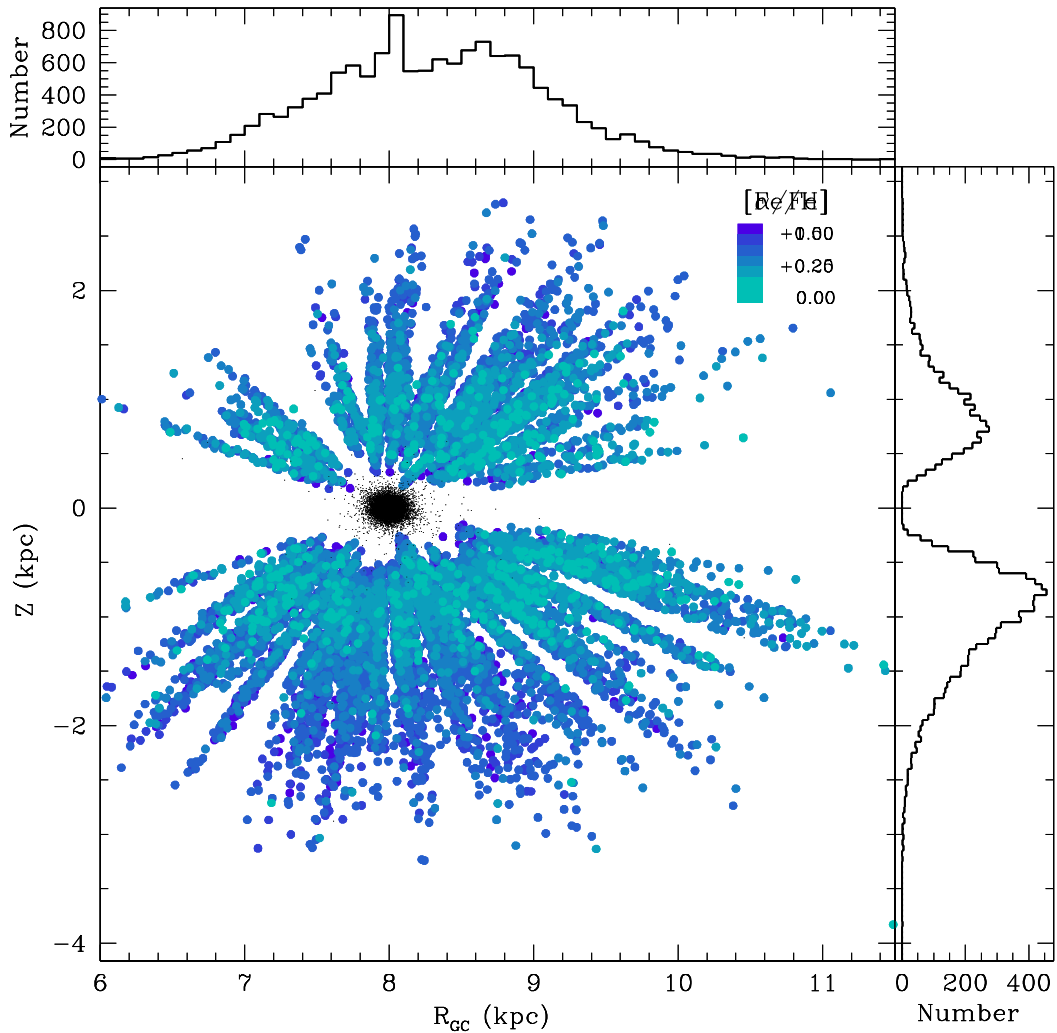}
\caption{The spatial coverage of our SEGUE G-dwarf sample. The top histogram shows the distribution of the sample in Galactocentric radius; the right-hand histogram shows the distribution in height above and below the plane of the Galaxy. The different colors for each point represent the [$\alpha$/Fe] value, as estimated by the SSPP. Darker points are more $\alpha$-enhanced, and tend to be farther out in $|Z|$. For comparison, the small black points at the center show the volume coverage of the Geneva-Copenhagen Survey \citep{jorgensen00, nordstrom04, holmberg07, holmberg09, casagrande11}.  
}
\label{fig:spatial_alpha}
\end{center}
\end{figure}

Figure\,\ref{fig:spatial_alpha} shows the spatial coverage of our entire G-dwarf sample in Galactocentric radius projected onto the plane, $R$, and height above the plane, $Z$. The Sun's position is assumed to be at ($R$, $Z$) of (8.0, 0.0) kpc \citep{bovy09}. Whereas previous analyses of the vertical disk gradient have been limited in coverage, the SEGUE sample covers stars both close to and far from the plane of the Galaxy, including stars from both the proposed thin- and thick-disk components. This makes the SEGUE sample ideal for constraining the disk's vertical structure and examining the interconnectivity of the two proposed disk components. 

\subsection{Correcting for Metallicity Biases} 
\label{sec:selection_biases}

Due to the survey design and target-selection criteria of SEGUE, our G-dwarf sample, which is defined by color and magnitude cuts, is biased towards metal-poor stars \citep{schlesinger12}. Specifically, 
\begin{itemize} 
\item Other SEGUE target categories that are biased towards metal-poor stars overlap the color range of the SEGUE G dwarfs, biasing
  the color-selected sample in metallicity.
\item SEGUE has the same limited number of spectroscopic fibers for each line of sight, regardless of its stellar density. Therefore, for lines of sight at large $|b|$, the spectroscopic sample is a higher fraction of the available photometric stars than for lines of sight at lower Galactic latitudes. 
\item The G-dwarf color cut selects a range of stellar masses, and thus a subset of the mass function, that varies with
  metallicity.
\end{itemize}

In \citet{schlesinger12}, we utilised SDSS photometry from DR7 \citep{abazajian09} to develop three weights for each target that account for these biases, such that the corrected sample reflects the true underlying structure of the Milky Way probed. Each individual spectroscopic target reflects the properties of similar photometric targets in SDSS and is weighted to reflect the number of similar stars. Since this work, the SDSS photometry has been improved, and the SSPP has been updated. Consequently, our stellar sample has changed. Using the algorithms described in \citet{schlesinger12}, we have recalculated the target-type, $r$-magnitude, and mass-function weights for our DR9 SEGUE G-dwarf sample. Thus, our adjusted sample reflects the distribution of stars observed photometrically by SDSS along all of the SEGUE lines of sight. 

We have also improved our weighting algorithm to better deal with anomalously large weights. Previously, the faintest stars often had large $r$-magnitude weights, as there was a single spectroscopic observation representing many photometric targets; we trimmed these objects from the sample using a magnitude cut \citep{schlesinger12}. For the DR9 sample, we use a more stringent S/N criteria; this results in individual spectroscopic observations representing numerous photometric objects at a range of magnitudes, rather than just the faint end. To account for this problem, we remove any SEGUE star from the sample that is alone in a 0.5 $r$-magnitude bin. This action removes less than 1\% of stars from the sample. 

Although many previous analyses have used the SEGUE G-dwarf sample, their target selection is different than our own. For example, \citet{lee11b} select stars that are {\it targeted} by SEGUE as G dwarfs, whereas we use all stars that meet G-dwarf criteria, even if they are were assigned SEGUE fibers for a different reason. By selecting those stars that were targeted only as G-dwarfs, the sample will be biased {\it against} metal-poor stars. Due to the prioritisation scheme for SEGUE targets, a star that fulfils the color cuts of both a ``G-dwarf star" and ``metal-poor star" is more likely to receive a fiber as a member of the latter category. In contrast, we include all stars that meet the color and magnitude criteria, regardless of why they were targeted, making the sample more complete. The analysis of \citet{carrell12} is similarly biased against metal-poor stars, as they select stars that meet {\it only} the SEGUE selection criteria of F, G, or K dwarfs, rather than stars that fulfil multiple criteria. Consequently, \citet{carrell12} will under-represent ÓinterestingÓ F, G, and K targets, such as low-metallicity stars and other overlapping categories. This bias is particularly serious for the SEGUE K-dwarf category, which is allotted few SEGUE fibers \citep{schlesinger12}.

\subsection{Subsamples}
\label{sec:subsamples}

As discussed earlier, the Milky Way is often described as having a thin- and thick-disk component. Previous analyses of the disk chemistry divided their samples geometrically, kinematically, and/or chemically in order to examine these two structures. In this work, we have designed a series of different subsamples to examine the disk as a whole and in chemical subsets. In addition, we recreate the divisions of previous works for a useful comparison. 

Although we have around 40,000 G-dwarf stars in our SEGUE sample, our individual subsamples consist of approximately 1000 stars each. In contrast to many previous analyses, we limit each subsample in distance such that it is volume-complete (\S\,\ref{sec:vol_complete}). This approach dramatically decreases the number of spectroscopic targets for each subsample. However, our weighting scheme (\S\,\ref{sec:selection_biases}) adjusts our sampling such that each spectroscopic target represents many photometric targets \citep{schlesinger12}. For example, our volume-complete Full Sample contains 1434 spectroscopic targets but reflects the properties of approximately 13,360 photometric targets.

\subsubsection{Geometric Subsamples}
\label{sec:geom_break} 

\begin{figure}[htbp]
\begin{center}
\includegraphics[width=\textwidth]{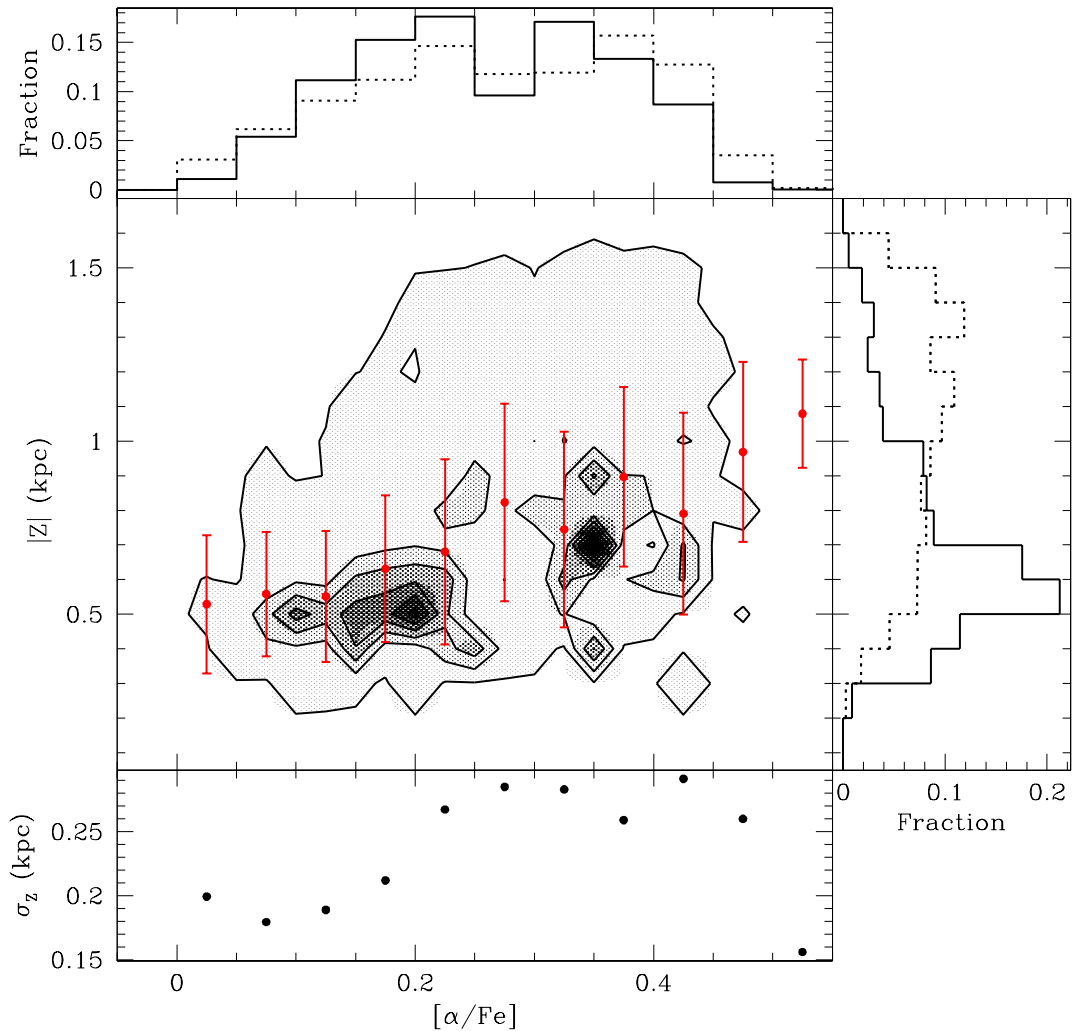}
\caption{The weighted distribution of our G-dwarf stars in [$\alpha$/Fe] vs. $|Z|$ space for the volume-complete Full Sample. The contours are in increments of 100 stars. The solid black histograms show the distribution in [$\alpha$/Fe] (top) and $|Z|$ (right) for the weighted sample. The dotted lines are the distribution for the raw G-dwarf sample. These distributions demonstrate the importance of accounting for target-selection biases in the SEGUE sample. In particular, the distribution in $|Z|$ becomes much more skewed to low $|Z|$ values. The red points represent the mean $|Z|$ value for a 0.05 dex bin in [$\alpha$/Fe], with the error bars showing the 1$\sigma$ range. The mean $|Z|$ value increases with increasing [$\alpha$/Fe]. Similarly, $\sigma_{|Z|}$ typically increases as [$\alpha$/Fe] increases (bottom panel).  
}
\label{fig:afezscatter}
\end{center}
\end{figure}

The stellar number density with respect to height above the plane of the Milky Way disk is best fit by an exponential thin- and thick-disk component, with scale heights of around 300 pc and 900 pc, respectively \citep{gilmore83, juric08}. Due to the differences in scale height, we expect few thin-disk stars at large heights above the Galactic plane. To isolate the thick-disk population, previous works limited their samples to stars with $|Z|\geq$1.0 kpc \citep{allendeprieto06, katz11, kordopatis11, chen11, carrell12}. We see a drop-off in star counts above heights of 1.0 kpc (Figure\,\ref{fig:afezscatter}). Although it is unclear that a geometric division can effectively separate the disk components, not to mention whether the two are in fact separable, for the sake of comparison with literature results, we define a subsample that consists of all G-dwarf stars with $|Z|$ greater than 1 kpc. 

\subsubsection{Chemical Subsamples}
\label{sec:chem_subsamples}

\begin{figure}[htbp]
\begin{center}
\includegraphics[width=\textwidth]{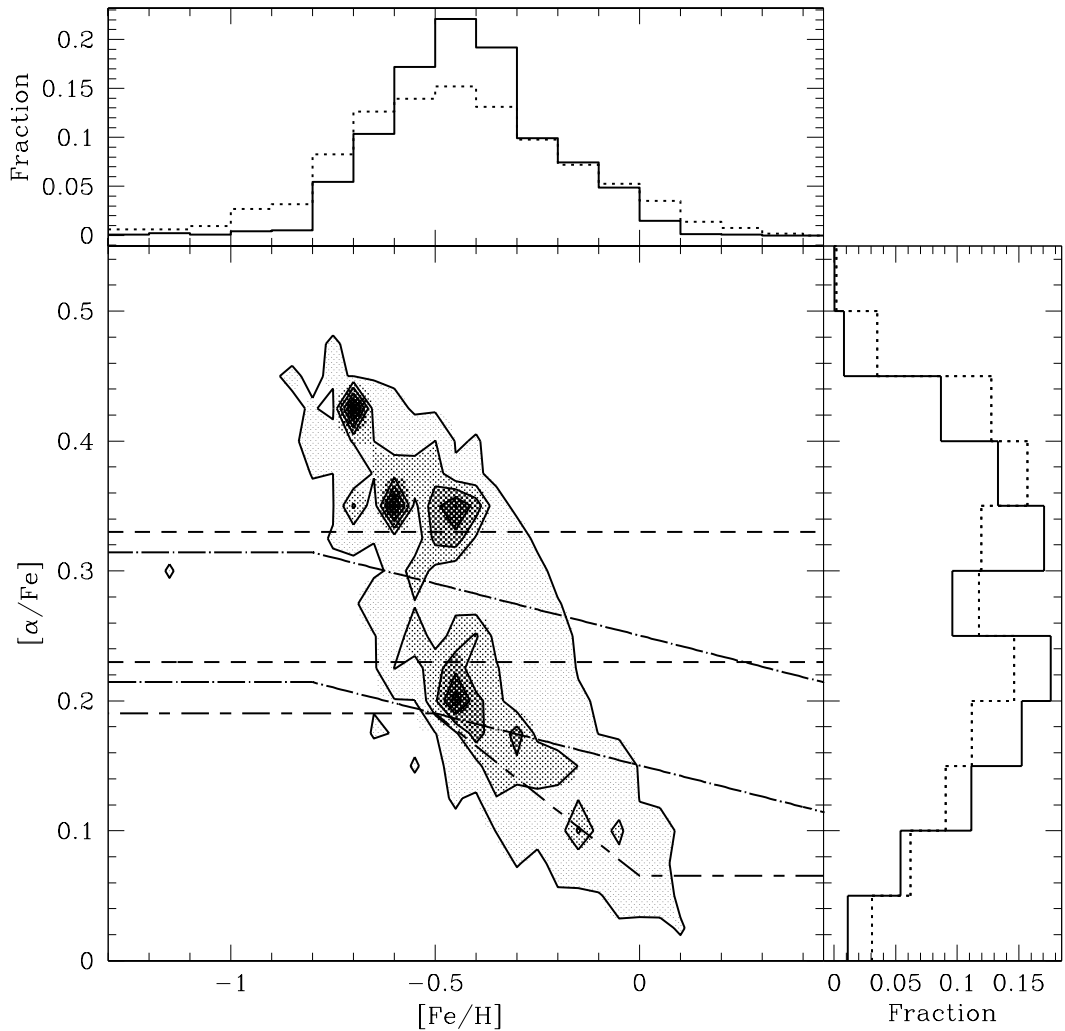}
\caption{The weighted distribution of our G-dwarf stars in [$\alpha$/Fe] vs. [Fe/H] space for the volume-complete Full Sample. The solid black histograms are the distribution in [Fe/H] (top) and [$\alpha$/Fe] (right) for the weighted sample. The dotted lines are the distribution for the raw G-dwarf sample. The contours are in increments of 150. The dot-dash line represents the ``thin" and ``thick"  disk separation from \citet{lee11b}. The short-dash lines show our own chemical separation, with $\alpha$-rich stars above [$\alpha$/Fe] $= +0.33$ and $\alpha$-poor below [$\alpha$/Fe] $= +0.23$. Finally, the long-short dash line is the thin/thick disk chemical separation from \citet{adibekyan}.  
}
\label{fig:afefeh_scatter}
\end{center}
\end{figure}

Figure\,\ref{fig:afefeh_scatter} shows the distribution of our weighted SEGUE G-dwarf sample in chemical space. As described earlier, and discussed at length in \citet{schlesinger12}, when we correct for the selection biases in SEGUE, the distribution shifts more to metal-rich and $\alpha$-poor values. 

Chemical abundance is oftentimes associated with age (\citealt{wg88}; \citealt{bovy12b}; \citealt{haywood13}), as [$\alpha$/Fe] and [Fe/H] proportions relate to the enrichment from SNe Type II and Ia, which contribute to the interstellar medium on different timescales (e.g., Maoz et al. 2011). Thus, examination of the vertical metallicity gradient with respect to [$\alpha$/Fe] provides an indication of how disk structure varies for different epochs of star formation. We divide our sample into 0.1 dex bins of [$\alpha$/Fe] to examine how the vertical metallicity gradient varies with $\alpha$-enhancement. There are a few stars with [$\alpha$/Fe]$> +0.5$, but not a sufficient number to populate their own bin, so we combine the two highest [$\alpha$/Fe] bins into a single bin over 0.2 dex in [$\alpha$/Fe]. Previous work by \citet{bovy12a, bovy12b} divided the SEGUE G-dwarf sample into mono-abundance populations in [$\alpha$/Fe] and [Fe/H]. In contrast, we do not limit our sample to 0.1 dex in [Fe/H] because we want to examine how metallicity varies over the volume of the disk.  

\citet{lee11b} detected bimodality in the raw SEGUE G-dwarf sample; they used the separation in [$\alpha$/Fe] space to define an $\alpha$-rich and $\alpha$-poor sample, associated with the thick and thin disk respectively (see their Figure 2). There are some important differences between our analysis and that of \citet{lee11b}. First, our sample of G dwarfs is different than that of \citet{lee11b}. As noted in \S\,\ref{sec:selection_biases}, they select only stars targeted as G dwarfs by SEGUE, biasing their sample against low-metallicity stars. Secondly, we trim our sample to be volume-complete, as discussed at length in \S\,\ref{sec:vol_complete}. While these actions do not necessarily affect the distribution in chemical space, they are important to consider when calculating the vertical metallicity gradients. 

While we observe [$\alpha$/Fe] bimodality in our bias-corrected sample, the thin/thick disk cuts of \citet{lee11b} divide the metal-rich, $\alpha$-poor peak of both our raw and weighted distributions, rather than clearly separating the sample into two $\alpha$-populations (see Figure\,\ref{fig:afefeh_scatter}). The \citet{lee11b} [$\alpha$/Fe] cuts are not well-suited to our observed bimodal distribution. This result is likely due to changes in the SSPP between DR8 and DR9. Examining the distribution of our sample in chemical space, we define our own thin/thick disk chemical subsamples based upon the observed bimodality at  [$\alpha$/Fe] $=$ 0.28. We define all stars with [$\alpha$/Fe]$ > +$0.33 as our $\alpha$-rich subsample, associated with the thick disk. Stars with [$\alpha$/Fe]$ < +$0.23 are our $\alpha$-poor subsample, associated with the thin disk. Similar to \citet{lee11b}, we also include a ``buffer" of 0.1 dex to prevent the two subpopulations from contaminating one another due to uncertainties in the SSPP [$\alpha$/Fe] estimates. For the sake of comparison with the literature, we also determine the vertical metallicity gradients of the \citet{lee11b} chemical subsamples. 

In contrast to \citet{lee11b}, \citet{bovy12a, bovy12b} found no bimodality in [$\alpha$/Fe] for an adjusted sample of SEGUE G-dwarf stars. Similar to our technique, they correct for the SEGUE selection function and use all stars that fulfil the G-dwarf color cut, regardless of why it was targeted by SEGUE. However, they adjust the sample further, by using stellar-population modelling and an assumed star-formation history to account for the limited lines of sight in SEGUE. As shown in Figure\,\ref{fig:spatial_alpha_aitoff}, the SEGUE survey probes 144 lines of sight. While we correct our spectroscopic sample such that it reflects the underlying populations for each of these lines of sight, \citet{bovy12a, bovy12b} seek to model the disk as a whole, using assumptions about the Galactic disk structure to scale up the SEGUE lines of sight. As much of SEGUE focused on lines of sight at large Galactic latitude, it contains more metal-poor, "thick disk" stars than a sample that uniformly samples the disk. Thus, when they adjust their sample using Galaxy models, they scale down the metal-poor stars and scale up the metal-rich, finding a smooth decrease in number as [Fe/H] decreases. Such an approach can be strongly impacted by uncertainties in the underlying Galaxy model; thus, we limit ourselves to analysing the underlying populations along the SEGUE lines of sight. While we observe bimodality in our SEGUE lines of sight after accounting for the local selection function, \citet{bovy12a, bovy12b} find no bimodality when they adjust the sample to reflect chemistry beyond the SEGUE sampling. 

Finally, recent work by \citet{adibekyan} on the High Accuracy Radial Velocity Planet Searcher (HARPS; \citealt{mayor03}) F-, G-, and K-dwarf sample identified a low-density region in the distribution of [$\alpha$/Fe] vs. [Fe/H]. Figure\,\ref{fig:afefeh_scatter} displays their separation superposed on our distribution. Due to the low resolution of SEGUE spectra, any gap will be smeared out in our sample. Additionally, in contrast to HARPS, SEGUE is dominated by stars farther from the plane of the Galaxy. The separable thin-disk population identified in HARPS is not well-sampled by SEGUE, and, consequently, will be difficult to detect. Thus, it is not surprising that their thin/thick separation is not observed in our sample of stars.

\subsection{Volume Completeness} 
\label{sec:vol_complete}

Previous works examined the vertical metallicity gradient over large distance ranges that are {\it not} volume-complete. For a given $(g-r)_0$ color, metal-rich stars are brighter than metal-poor stars. Thus, due to issues of survey completeness and saturation limits, the sample will be biased towards metal-rich stars at large distances and toward metal-poor at small distances. We specify distance limits for each of our subsamples such that they are volume-complete. 

Our subsamples are typically defined in terms of [$\alpha$/Fe]. For each of these, we examine the range of $r_0$ and [Fe/H] to determine an appropriate volume-complete distance range. The bright magnitude limit is set to $r = $15, the saturation limit for SEGUE. We also institute a faint magnitude limit; if we do not trim the sample in magnitude, we occasionally include quite faint stars which result in an unreal distance range (e.g., the faint limit is closer than the bright limit). For the faint magnitude limit, we select the $r_0$ for which 85\% of the sample is brighter, $r_0 = $17.13, following the methodology of \citet{schlesinger12}. 

The most metal-poor stars with the least amount of $\alpha$-enhancement define the faint distance limit; the most metal-rich stars with the maximum amount of [$\alpha$/Fe] define the bright distance limit. Some of our subsamples have outliers in [Fe/H], which can lead to an unreal distance range, as with the magnitude range. To ensure that our distance limits reflect the overall parameters of the sample, rather than being skewed by outliers, we remove any stars in the extreme 5\% of the distribution for each subsample. 

For each of our defined subsamples, we generate 10 Gyr isochrones for the two possible chemical extremes of each subsample using the Dartmouth Isochrone Generator 2012 \citep{dartmouth}; these correspond to the faint and bright distance limits. We then extract $M_{r}$ at $(g-r)$ of 0.48 and 0.55. Table\,\ref{tab:dist_limits} lists the properties of each subsample and its associated distance limits. 

\citet{schlesinger12} restrict their full G-dwarf sample to distances between 1.59 and 2.29 kpc for volume-completeness. Our distance limits for the Full Sample are much more limited, between 1.447 and 1.614 kpc. We want to ensure that all of our stars have [$\alpha$/Fe] measurements, which requires a $S/N >$ 30, stricter than the $S/N > $10 constraint of \citet{schlesinger12}. Not only does this $S/N$ criteria limit our sample size, it also affects our faint magnitude limit. Whereas the sample in \citet{schlesinger12} has stars as faint as $r_0 = $18.45, our limitation to $r_0 = $17.13 greatly decreases our maximum distance limit, reducing the number of stars in each of our chemical subsamples. 

\begin{deluxetable}{cccc}
\tabletypesize{\small}
\tablewidth{0pt}
\tablecaption{Distance and Chemical Limits for Different Subsamples \label{tab:dist_limits}}
\tablehead{
\colhead{Subsample} & \colhead{[Fe/H]} & \colhead{Distance (kpc)} & \colhead{[$\alpha$/Fe]} }
\startdata
Full Sample                      &  $-1.00$ to $-0.04$ & 1.447 to 1.614 & \dots  \\
$\alpha$-rich         	      &  $-1.00$ to $-0.45$     & 1.182 to 1.690 & $>$ +0.33 \\
$\alpha$-poor       	      &  $-0.53$ to $0.09$      & 1.273 to 2.034 & $<$ +0.23 \\
Lee $\alpha$-rich            &  $-1.00$ to $-0.42$  & 1.232 to 1.690 &  +0.28 to +0.54 \\
Lee $\alpha$-poor           &  $-0.49$ to $0.15$   & 1.337 to 2.075 & +0.01 to +0.18 \\ 
$\alpha$-bin 1                  &  $-0.24$ to $0.24$   & 1.351 to 2.547 & +0.0 to +0.1 \\
$\alpha$-bin 2                  &  $-0.50$ to $-0.05$  & 1.189 to 2.120 & +0.1 to +0.2 \\
$\alpha$-bin 3     	       &  $-0.68$ to $-0.22$     & 1.348 to 1.946 & +0.2 to +0.3 \\
$\alpha$-bin 4        	       &  $-1.00$ to $-0.38$     & 1.179 to 1.690 & +0.3 to +0.4 \\
$\alpha$-bin 5                 &  $-1.00$ to $-0.59$   & 1.029 to 1.730 & +0.4 to +0.6 \\
\enddata 
\tablecomments{
Our subsamples and their [Fe/H] and [$\alpha$/Fe] coverage. Using these values and a $r_0$ magnitude range from 15.00 to 17.13, we use 10 Gyr Dartmouth isochrones to determine the volume-complete distance range for each subsample. We trim each subsample to remove stars in the extreme 5\% of the [Fe/H] distribution, in order to avoid outliers skewing the vertical metallicity gradient and leading to an unreal volume-complete distance range. }
\end{deluxetable}

\section{Determining the Vertical Metallicity Gradient} 
\label{sec: vmg}

For each of our defined subsamples, we now examine how the median [Fe/H] varies with respect to height above the Galactic plane ($|Z|$). We sort each subsample by $|Z|$ and scale individual stars by their associated target-selection weights. We then apply a boxcar-smoothing technique to each subsample. Specifically, we determine the median [Fe/H] and $|Z|$ of the 10\% of the subsample at the lowest $|Z|$; this is the first point in our vertical metallicity gradient. We then step through the weighted sample in 100-star increments, determining the median [Fe/H] and $|Z|$. 

\subsection{Effects of Accounting for Target-selection Biases} 

\begin{figure}[htbp]
\begin{center}
\includegraphics[width=\textwidth]{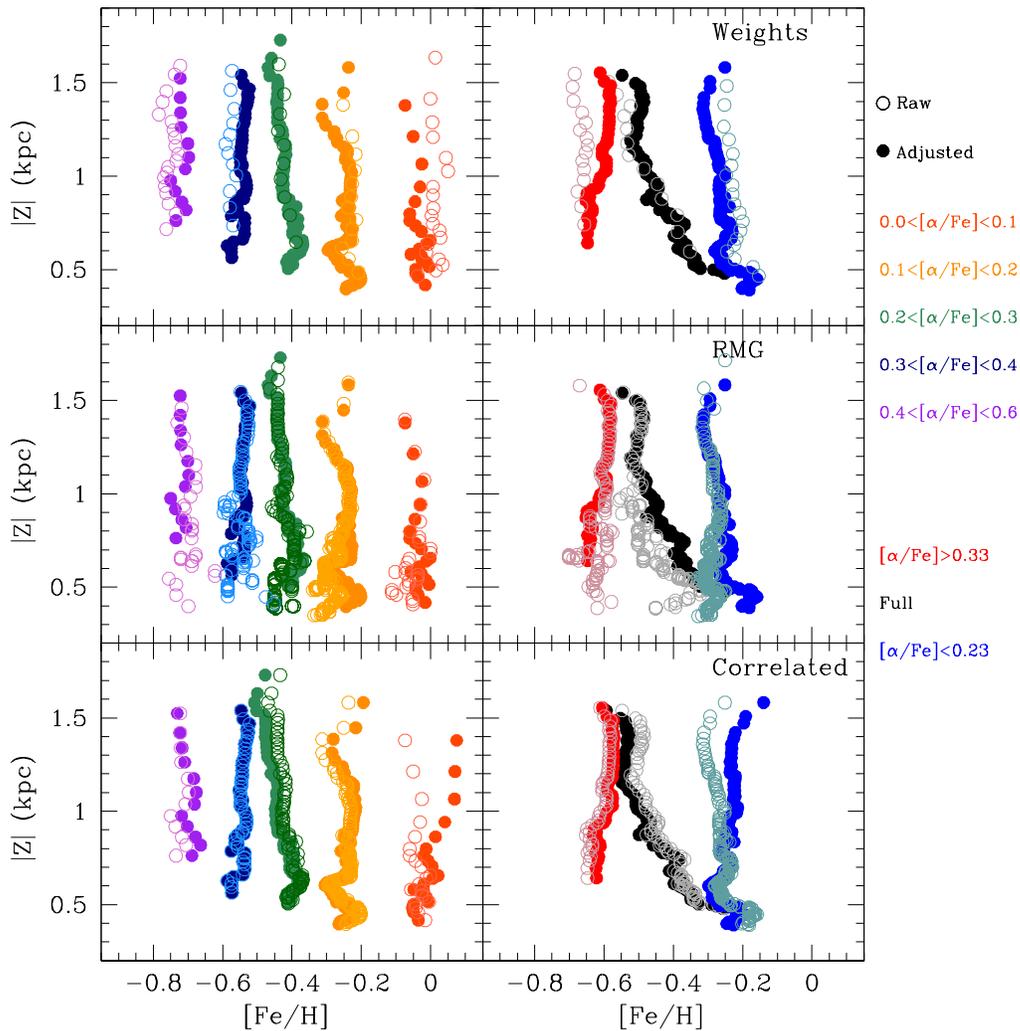}
\caption{Changes in the vertical metallicity gradients due to different sample adjustments. The original gradient is represented by the open points; the adjusted is shown by the solid points. Each color represents a different chemical subsample, as listed on the right-hand side of the figure.
\emph{Top:} Changes due to accounting for SEGUE target selection, which is biased towards metal-poor stars. This correction affects both the zero-point and the slope of our vertical metallicity gradients. Changes in slope are listed in Table\,\ref{tab:correlated_errors}.
\emph{Middle:} The changes in the gradients due to corrections for the radial metallicity gradient using the slopes determined by \citet{cheng12b}. 
\emph{Bottom:} The open points are the original weighted vertical metallicity gradient, with a correction for the radial metallicity gradient. The solid points are corrected for the correlated errors in [Fe/H] and distance. The change in slope is listed in Table\,\ref{tab:correlated_errors}. 
}
\label{fig:triple_whammy}
\end{center}
\end{figure}

As noted in \S\,\ref{sec:selection_biases}, the SEGUE G-dwarf sample is biased toward metal-poor stars due to the survey's target-selection algorithm. Figure\,\ref{fig:triple_whammy} demonstrates that accounting for these biases has a marked affect on both the slope and zero-point of the vertical metallicity gradients of the different subsamples. The slopes of the ``raw" samples are listed in Table\,\ref{tab:correlated_errors}. 

Although we have removed stars that are alone in their $r$ magnitude bin from the sample, some of the G-dwarf stars still have anomalously large target-selection weights. Typically, these are stars in magnitude bins that have a small number of spectroscopic targets and a large number of photometric targets. Due to our stringent cuts in spectral $S/N$, these heavily-weighted stars occur at a range of magnitudes. These anomalous weights will induce large wiggles in our vertical metallicity gradient, especially at low $|Z|$ where the stellar density is high but the SEGUE sample size is small. For each subsample, we remove all stars with weights that lie 2$\sigma$ from the mean weight. This procedure tends to remove more metal-rich stars, as these are more likely to be in high-density regions close to the plane. Although it has a small effect on the intercept, removing these anomalous weights does not significantly change the measured vertical metallicity gradient; rather it serves to smooth the structure with respect to $|Z|$. Most of the gradients measured for the untrimmed subsamples lie within 1$\sigma$ of the values of the trimmed sample; all are within 2$\sigma$. 

\subsection{Correcting for the Radial Metallicity Gradient} 

Our stellar sample ranges from 6 to 11 kpc in Galactocentric radius (Figure\,\ref{fig:spatial_alpha}); over a volume-complete range, it is limited to 6.7 to 9.5 kpc. Thus, our sample is not well-suited to determining the radial metallicity gradient. However, our sample is still affected by the Galactic radial metallicity gradient, which in turn will affect our estimates of the vertical metallicity gradient for different subsamples. 

\citet{cheng12a} measured radial metallicity gradients for an unbiased low-latitude sample of SEGUE main-sequence turnoff stars, ranging from 6 to 16 kpc in Galactocentric radius. They also divided the sample into bins of [$\alpha$/Fe] and $|Z|$, finding that stars with [$\alpha$/Fe]$ > +$0.2 showed a flat radial metallicity gradient at all $|Z|$. In contrast, the radial metallicity gradient of stars with [$\alpha$/Fe]$ < +$0.2 flattened with increasing $|Z|$, similar to the sample as a whole.

Based on each individual star's chemistry and Galactocentric radius, we adjust the stellar [Fe/H] using the slopes from \citet{cheng12b} such that they reflect the metallicity at the solar radius (see Figure\,\ref{fig:triple_whammy}). Other analyses have estimated the radial metallicity gradient (e.g., \citealt{maciel2010}, \citealt{boeche2013}, \citealt{hayden13}), but we adopt the values from \citet{cheng12b} as they divide their sample in [$\alpha$/Fe] and cover a similar range of $|Z|$ with stars observed \emph{in situ}. Figure\,\ref{fig:triple_whammy} shows the changes in our vertical metallicity gradient when we remove the effect of radial metallicity structure is removed. The SEGUE G-dwarf sample is typically at larger $|Z|$, where the radial gradient is minimal. The largest change in [Fe/H], $\approx$0.05 dex, occurs at low $|Z|$ and primarily affects the $\alpha$-poor samples, which lie close to the Galactic plane. 

\subsection{Correlated Uncertainties: [Fe/H] and Distance}
\label{sec:corr_errors}

In addition to corrections for SEGUE target selection and the radial metallicity gradient, we must also examine the systematic uncertainties in the vertical metallicity gradient stemming from correlated errors in estimates of [Fe/H] and distance. An underestimate in [Fe/H] will lead to an underestimated distance, and vice versa, moving targets into different bins of R, $|Z|$, and [Fe/H] \citep{schlesinger12}, which can induce an artificial slope in the derived vertical metallicity gradient.  

To determine how correlated errors affect our vertical metallicity gradients, we must analyse a sample where we know the ``true" underlying behavior. Thus, we simulate each SEGUE line of sight in our sample using the chemical and dynamical Galaxy model of \citet{schonrich09a, schonrich09b}. This model is notable for including radial migration in addition to simulating chemical evolution. Furthermore, it provides both [Fe/H] and [$\alpha$/Fe] values. \citet{schlesinger12} found that this model did not accurately reflect the vertical metallicity structure of their observed sample, becoming more metal rich with increasing $|Z|$, whereas the SEGUE sample becomes more metal poor. Although the model does not recreate the observed distributions, the absolute metallicity structure is not of major consequence for the purpose of this error analysis. We are simply examining at the variation in the sample caused by correlated errors. 

Applying SEGUE and subsample selection criteria to the model, we randomly assign each star an uncertainty in metallicity, based on the expected SSPP errors. We then determine the resulting change in distance for each model star, and shift its distance, $R$, and $Z$ accordingly. We run this procedure 20 times to produce 20 iterations of the simulated SEGUE fields. For each iteration, we perform a least-squares fit to the vertical metallicity gradient with the correlated uncertainties in [Fe/H] and distance and compare the median slope to the true model vertical metallicity gradient for each subsample. This calculation reveals the degree to which correlated errors may change our measured gradients. We also examine the variation in the metallicity gradient with correlated errors to estimate the 1$\sigma$ uncertainty in this correction. For the vertical metallicity gradient of each subsample, we subtract the estimated induced slope from our value of the weighted vertical metallicity gradient. Table\,\ref{tab:correlated_errors} lists the corrections, and their associated uncertainty, for correlated errors for each subsample. For a given $\Delta$[Fe/H], the magnitude difference between metal-rich isochrones is larger than that between metal-poor for the main sequence. Consequently, the change in estimated distance due to $\Delta$[Fe/H] will be larger at the metal-rich end, leading to a larger uncertainty from correlated errors in the vertical metallicity gradient. Figure\,\ref{fig:triple_whammy} presents the change in the vertical metallicity gradients when we account for correlated errors; these corrections vary for different subsamples, but are typically around $\pm$0.08 dex kpc$^{-1}$. 

\subsection{Uncertainties in the Metallicity Gradient} 
\label{sec:grad_uncertainties} 

To estimate the uncertainties in our vertical metallicity gradients, we combine information from a bootstrap analysis (with replacement) and a Monte Carlo analysis. Using these two techniques, we quantify the uncertainty in [Fe/H] at each $|Z|$ and the error on the measured slope for each chemical subsample. To determine the total uncertainty in median [Fe/H] at each height, we combine the errors from the following sources: sample selection, photometry, extinction, random and systematic distance, [$\alpha$/Fe], radial metallicity gradient correction, binarity, and $\log g$. We discuss our methodology in detail in Appendix\,\ref{appendix}. Uncertainties from our bootstrap analysis, related to variation in sample selection, contribute the largest error at each $|Z|$ for the different subsamples, around 0.06 dex. Close to the plane of the Galaxy, there are fewer SEGUE stars and a high stellar number density. Consequently, our uncertainties in [Fe/H] are larger at low $|Z|$. 

We also determine the variation in slope due to the individual sources of error in our sample, namely sample selection, $\log g$, [$\alpha$/Fe], distance, photometry, reddening, undetected binarity,  and radial metallicity gradient corrections (Table\,\ref{tab:correlated_errors}). The results from our bootstrap analysis produce the largest shifts in slope, around $\pm$0.04 dex kpc$^{-1}$, indicating that much of the uncertainty is driven by sample selection.

\begin{landscape}
\begin{deluxetable}{ccccccccccccccc@{}lc@{}l}
\tabletypesize{\tiny}
\tablewidth{0pt}
\setlength{\tabcolsep}{0.02in} 
\tablecaption{Vertical Metallicity Gradient for Different Subsamples \label{tab:correlated_errors}}
\tablehead{
\multirow{2}{*}{Sample} & \multirow{2}{*}{Number} & \colhead{Weighted} &\multirow{2}{*}{$\frac{d[Fe/H]}{d|Z|}_{raw}$} & \multirow{2}{*}{$\frac{d[Fe/H]}{d|Z|}_{wt}$} & \multicolumn{9}{c}{$\sigma$} & \multirow{2}{*}{$\Delta \frac{d[Fe/H]}{d|Z|}_{CE}$}  & & \multirow{2}{*}{$\frac{d[Fe/H]}{d|Z|}$} &  \\
\colhead{} & \colhead{} & \colhead{Number} &\colhead{} & \colhead{} & \colhead{Bootstrap} &  \colhead{$\log g$} &  \colhead{[$\alpha$/Fe]} &   \colhead{D$_{rand}$} &  \colhead{D$_{syst}$} &  \colhead{Phot} &  \colhead{Redden} &  \colhead{Binary} & \colhead{RGC} & \colhead{} & \colhead{} & \colhead{} & \colhead{}  \\
\colhead{(1)} &
\colhead{(2)} &
\colhead{(3)} &
\colhead{(4)} &
\colhead{(5)} &
\colhead{(6)} &
\colhead{(7)} &
\colhead{(8)} &
\colhead{(9)} &
\colhead{(10)} &
\colhead{(11)} &
\colhead{(12)} &
\colhead{(13)} &
\colhead{(14)} &
\colhead{(15)} &
\colhead{(16)} &
\colhead{(17)} &
\colhead{(18)} 
}
\startdata
\multirow{2}{*}{Full Sample} 	 & \multirow{2}{*}{1434} 	 & \multirow{2}{*}{13359} 	 & \multirow{2}{*}{$-$0.252} 	 & \multirow{2}{*}{$-$0.206} 	 & 0.000 	 & 0.002 	 & 0.003 	 & 0.006 	 & 0.009 	 &0.002 	 & 0.000 	 & 0.004 	 & 0.031 	 & \multirow{2}{*}{$+$0.037} 	 & 0.020 	 & \multirow{2}{*}{$-$0.243} 	 & 0.039 \\ 
 	 & 	 & 	 & 	 & 	 & 0.042 	 & 0.004 	 & 0.001 	 & 0.009 	 & 0.006 	 & 0.002 	 & 0.006 	 & 0.002 	 & $-$0.018 	 & 	 & 0.022 	 & 	 & 0.053 \\ 
\\
\multirow{2}{*}{$|Z|>$1 kpc} 	 & \multirow{2}{*}{806} 	 & \multirow{2}{*}{2080} 	 & \multirow{2}{*}{$-$0.133} 	 & \multirow{2}{*}{$-$0.102} 	 & 0.096 	 & 0.032 	 & 0.033 	 & 0.082 	 & 0.112 	 &0.022 	 & 0.018 	 & 0.026 	 & 0.012 	 & \multirow{2}{*}{$+$0.131} 	 & 0.107 	 & \multirow{2}{*}{$-$0.233} 	 & 0.209 \\ 
 	 & 	 & 	 & 	 & 	 & 0.047 	 & 0.018 	 & 0.010 	 & 0.000 	 & 0.000 	 & 0.012 	 & 0.015 	 & 0.016 	 & 0.003 	 & 	 & 0.137 	 & 	 & 0.149 \\ 
\\
\multirow{2}{*}{$\alpha$-rich} 	 & \multirow{2}{*}{1614} 	 & \multirow{2}{*}{8996} 	 & \multirow{2}{*}{$-$0.029} 	 & \multirow{2}{*}{$+$0.083} 	 & 0.038 	 & 0.006 	 & 0.002 	 & 0.008 	 & 0.004 	 &0.002 	 & 0.006 	 & 0.001 	 & 0.002 	 & \multirow{2}{*}{$+$0.021} 	 & 0.024 	 & \multirow{2}{*}{$+$0.063} 	 & 0.047 \\ 
 	 & 	 & 	 & 	 & 	 & 0.024 	 & 0.003 	 & 0.005 	 & 0.011 	 & 0.007 	 & 0.005 	 & 0.003 	 & 0.006 	 & 0.003 	 & 	 & 0.012 	 & 	 & 0.032 \\ 
\\
\multirow{2}{*}{$\alpha$-poor} 	 & \multirow{2}{*}{2042} 	 & \multirow{2}{*}{24550} 	 & \multirow{2}{*}{$-$0.063} 	 & \multirow{2}{*}{$-$0.094} 	 & 0.020 	 & 0.003 	 & 0.001 	 & 0.001 	 & 0.000 	 &0.001 	 & 0.002 	 & 0.000 	 & 0.035 	 & \multirow{2}{*}{$-$0.132} 	 & 0.014 	 & \multirow{2}{*}{$+$0.038} 	 & 0.043 \\ 
 	 & 	 & 	 & 	 & 	 & 0.018 	 & 0.002 	 & 0.003 	 & 0.004 	 & 0.004 	 & 0.001 	 & 0.002 	 & 0.003 	 & $-$0.029 	 & 	 & 0.011 	 & 	 & 0.037 \\ 
\\
\multirow{2}{*}{Lee $\alpha$-rich} 	 & \multirow{2}{*}{1795} 	 & \multirow{2}{*}{12008} 	 & \multirow{2}{*}{$-$0.032} 	 & \multirow{2}{*}{$+$0.025} 	 & 0.027 	 & 0.003 	 & 0.001 	 & 0.010 	 & 0.003 	 &0.001 	 & 0.003 	 & 0.007 	 & $-$0.009 	 & \multirow{2}{*}{$+$0.045} 	 & 0.016 	 & \multirow{2}{*}{$-$0.020} 	 & 0.035 \\ 
 	 & 	 & 	 & 	 & 	 & 0.024 	 & 0.006 	 & 0.007 	 & 0.009 	 & 0.005 	 & 0.003 	 & 0.006 	 & 0.005 	 & 0.002 	 & 	 & 0.015 	 & 	 & 0.033 \\ 
\\
\multirow{2}{*}{Lee $\alpha$-poor} 	 & \multirow{2}{*}{1064} 	 & \multirow{2}{*}{12802} 	 & \multirow{2}{*}{$-$0.034} 	 & \multirow{2}{*}{$-$0.078} 	 & 0.059 	 & 0.003 	 & 0.000 	 & 0.002 	 & 0.000 	 &0.000 	 & 0.001 	 & 0.000 	 & 0.041 	 & \multirow{2}{*}{$-$0.164} 	 & 0.016 	 & \multirow{2}{*}{$+$0.087} 	 & 0.074 \\ 
 	 & 	 & 	 & 	 & 	 & 0.015 	 & 0.002 	 & 0.009 	 & 0.006 	 & 0.005 	 & 0.003 	 & 0.003 	 & 0.004 	 & $-$0.054 	 & 	 & 0.015 	 & 	 & 0.059 \\ 
\\
\multirow{2}{*}{$\alpha$-bin 1} 	 & \multirow{2}{*}{445} 	 & \multirow{2}{*}{2874} 	 & \multirow{2}{*}{$+$0.009} 	 & \multirow{2}{*}{$-$0.038} 	 & 0.039 	 & 0.005 	 & 0.003 	 & 0.003 	 & 0.001 	 &0.002 	 & 0.002 	 & 0.002 	 & 0.024 	 & \multirow{2}{*}{$-$0.176} 	 & 0.011 	 & \multirow{2}{*}{$+$0.138} 	 & 0.048 \\ 
 	 & 	 & 	 & 	 & 	 & 0.037 	 & 0.000 	 & 0.004 	 & 0.003 	 & 0.003 	 & 0.002 	 & 0.003 	 & 0.002 	 & $-$0.025 	 & 	 & 0.023 	 & 	 & 0.051 \\ 
\\
\multirow{2}{*}{$\alpha$-bin 2} 	 & \multirow{2}{*}{1448} 	 & \multirow{2}{*}{18145} 	 & \multirow{2}{*}{$-$0.008} 	 & \multirow{2}{*}{$-$0.026} 	 & 0.015 	 & 0.006 	 & 0.005 	 & 0.004 	 & 0.001 	 &0.002 	 & 0.002 	 & 0.003 	 & 0.020 	 & \multirow{2}{*}{$-$0.053} 	 & 0.018 	 & \multirow{2}{*}{$+$0.027} 	 & 0.032 \\ 
 	 & 	 & 	 & 	 & 	 & 0.018 	 & 0.001 	 & 0.006 	 & 0.004 	 & 0.003 	 & 0.002 	 & 0.003 	 & 0.002 	 & $-$0.020 	 & 	 & 0.018 	 & 	 & 0.033 \\ 
\\
\multirow{2}{*}{$\alpha$-bin 3} 	 & \multirow{2}{*}{1095} 	 & \multirow{2}{*}{12996} 	 & \multirow{2}{*}{$-$0.041} 	 & \multirow{2}{*}{$-$0.063} 	 & 0.026 	 & 0.007 	 & 0.010 	 & 0.014 	 & 0.000 	 &0.010 	 & 0.021 	 & 0.004 	 & 0.022 	 & \multirow{2}{*}{$+$0.034} 	 & 0.014 	 & \multirow{2}{*}{$-$0.097} 	 & 0.048 \\ 
 	 & 	 & 	 & 	 & 	 & 0.018 	 & 0.010 	 & 0.015 	 & 0.012 	 & 0.016 	 & 0.009 	 & 0.001 	 & 0.012 	 & $-$0.022 	 & 	 & 0.016 	 & 	 & 0.045 \\ 
\\
\multirow{2}{*}{$\alpha$-bin 4} 	 & \multirow{2}{*}{1261} 	 & \multirow{2}{*}{9432} 	 & \multirow{2}{*}{$-$0.000} 	 & \multirow{2}{*}{$+$0.031} 	 & 0.023 	 & 0.007 	 & 0.000 	 & 0.017 	 & 0.007 	 &0.007 	 & 0.010 	 & 0.012 	 & $-$0.001 	 & \multirow{2}{*}{$+$0.004} 	 & 0.029 	 & \multirow{2}{*}{$+$0.027} 	 & 0.045 \\ 
 	 & 	 & 	 & 	 & 	 & 0.027 	 & 0.012 	 & 0.065 	 & 0.008 	 & 0.005 	 & 0.001 	 & 0.014 	 & 0.007 	 & $-$0.016 	 & 	 & 0.035 	 & 	 & 0.083 \\ 
\\
\multirow{2}{*}{$\alpha$-bin 5} 	 & \multirow{2}{*}{890} 	 & \multirow{2}{*}{3085} 	 & \multirow{2}{*}{$+$0.013} 	 & \multirow{2}{*}{$+$0.008} 	 & 0.061 	 & 0.012 	 & 0.022 	 & 0.012 	 & 0.005 	 &0.006 	 & 0.000 	 & 0.011 	 & 0.017 	 & \multirow{2}{*}{$+$0.072} 	 & 0.018 	 & \multirow{2}{*}{$-$0.064} 	 & 0.073 \\ 
 	 & 	 & 	 & 	 & 	 & 0.025 	 & 0.001 	 & 0.002 	 & 0.007 	 & 0.005 	 & 0.003 	 & 0.008 	 & 0.005 	 & $-$0.004 	 & 	 & 0.020 	 & 	 & 0.035 \\ 
\enddata 
\tablecomments{
The estimated vertical metallicity gradients for our different subsamples with their associated uncertainties. Each subsample uses the distance limits listed in Table\,\ref{tab:dist_limits}; we have also determined the gradients for all subsamples over the distance range specified for the Full Sample; these values agree within 1$\sigma$. Column (2) lists the number of stars that fall within each subsample; Column (3) is the weighted number of stars after we have corrected for SEGUE target selection. Column (4) is the gradient measured for the raw stellar subsample. Column (5) lists the slope when we have adjusted for target-selection weights and the radial metallicity gradient, but not correlated errors. The change from correlated errors is listed in Column (15). Column (6) through (14) list the uncertainty on the slope due to our different sources of uncertainty, as detailed in Appendix\,\ref{appendix}. Column (17) presents our final estimated vertical metallicity gradient for each subsample, with the total uncertainty in Column (18). }
\end{deluxetable}
\end{landscape}

\begin{landscape}
\begin{deluxetable}{ccccccccccccccc@{}lc@{}l}
\tabletypesize{\tiny}
\tablewidth{0pt}
\setlength{\tabcolsep}{0.02in} 
\tablecaption{Vertical Metallicity Gradient for Different Subsamples with 1.447$<$D$<$1.614 kpc \label{tab:correlated_errors_overlap}}
\tablehead{
\multirow{2}{*}{Sample} & \multirow{2}{*}{Number} & \colhead{Weighted} & \multirow{2}{*}{$\frac{d[Fe/H]}{d|Z|}_{raw}$} & \multirow{2}{*}{$\frac{d[Fe/H]}{d|Z|}_{wt}$} & \multicolumn{9}{c}{$\sigma$} & \multirow{2}{*}{$\Delta \frac{d[Fe/H]}{d|Z|}_{CE}$}  & & \multirow{2}{*}{$\frac{d[Fe/H]}{d|Z|}$} &  \\
\colhead{} & \colhead{} & \colhead{Number} & \colhead{} & \colhead{} & \colhead{Bootstrap} &  \colhead{$\log g$} &  \colhead{[$\alpha$/Fe]} &   \colhead{D$_{rand}$} &  \colhead{D$_{syst}$} &  \colhead{Phot} &  \colhead{Redden} &  \colhead{Binary} & \colhead{RGC} & \colhead{} & \colhead{} & \colhead{} & \colhead{}  \\
\colhead{(1)} &
\colhead{(2)} &
\colhead{(3)} &
\colhead{(4)} &
\colhead{(5)} &
\colhead{(6)} &
\colhead{(7)} &
\colhead{(8)} &
\colhead{(9)} &
\colhead{(10)} &
\colhead{(11)} &
\colhead{(12)} &
\colhead{(13)} &
\colhead{(14)} &
\colhead{(15)} &
\colhead{(16)} &
\colhead{(17)} &
\colhead{(18)} 
}
\startdata
\multirow{2}{*}{Full Sample} 	 & \multirow{2}{*}{1434} 	 & \multirow{2}{*}{13359} 	 & \multirow{2}{*}{$-$0.252} 	 & \multirow{2}{*}{$-$0.206} 	 & 0.000 	 & 0.002 	 & 0.003 	 & 0.006 	 & 0.009 	 &0.002 	 & 0.000 	 & 0.004 	 & 0.031 	 & \multirow{2}{*}{$+$0.037} 	 & 0.020 	 & \multirow{2}{*}{$-$0.243} 	 & 0.039 \\ 
 	 & 	 & 	 & 	 & 	 & 0.042 	 & 0.004 	 & 0.001 	 & 0.009 	 & 0.006 	 & 0.002 	 & 0.006 	 & 0.002 	 & $-$0.018 	 & 	 & 0.022 	 & 	 & 0.053 \\ 
\\
\multirow{2}{*}{$|Z|>$1 kpc} 	 & \multirow{2}{*}{806} 	 & \multirow{2}{*}{2080} 	 & \multirow{2}{*}{$-$0.133} 	 & \multirow{2}{*}{$-$0.102} 	 & 0.096 	 & 0.032 	 & 0.033 	 & 0.082 	 & 0.112 	 &0.022 	 & 0.018 	 & 0.026 	 & 0.012 	 & \multirow{2}{*}{$+$0.131} 	 & 0.107 	 & \multirow{2}{*}{$-$0.233} 	 & 0.209 \\ 
 	 & 	 & 	 & 	 & 	 & 0.047 	 & 0.018 	 & 0.010 	 & 0.000 	 & 0.000 	 & 0.012 	 & 0.015 	 & 0.016 	 & 0.003 	 & 	 & 0.137 	 & 	 & 0.149 \\ 
\\
\multirow{2}{*}{$\alpha$-rich} 	 & \multirow{2}{*}{528} 	 & \multirow{2}{*}{4067} 	 & \multirow{2}{*}{$-$0.002} 	 & \multirow{2}{*}{$+$0.128} 	 & 0.060 	 & 0.012 	 & 0.004 	 & 0.013 	 & 0.004 	 &0.003 	 & 0.001 	 & 0.007 	 & 0.013 	 & \multirow{2}{*}{$+$0.035} 	 & 0.014 	 & \multirow{2}{*}{$+$0.093} 	 & 0.066 \\ 
	 & 	 & 	 & 	 & 	 & 0.051 	 & 0.003 	 & 0.008 	 & 0.023 	 & 0.018 	 & 0.006 	 & 0.016 	 & 0.015 	 & $-$0.022 	 & 	 & 0.025 	 & 	 & 0.072 \\ 
\\
\multirow{2}{*}{$\alpha$-poor } 	 & \multirow{2}{*}{538} 	 & \multirow{2}{*}{5871} 	 & \multirow{2}{*}{$-$0.085} 	 & \multirow{2}{*}{$-$0.136} 	 & 0.048 	 & 0.006 	 & 0.001 	 & 0.002 	 & 0.005 	 &0.002 	 & 0.001 	 & 0.004 	 & 0.037 	 & \multirow{2}{*}{$-$0.125} 	 & 0.067 	 & \multirow{2}{*}{$-$0.011} 	 & 0.091 \\ 
 	 & 	 & 	 & 	 & 	 & 0.028 	 & 0.001 	 & 0.007 	 & 0.014 	 & 0.003 	 & 0.001 	 & 0.009 	 & 0.003 	 & $-$0.041 	 & 	 & 0.031 	 & 	 & 0.062 \\ 
\\
\multirow{2}{*}{Lee $\alpha$-rich} 	 & \multirow{2}{*}{648} 	 & \multirow{2}{*}{5422} 	 & \multirow{2}{*}{$-$0.071} 	 & \multirow{2}{*}{$+$0.014} 	 & 0.088 	 & 0.014 	 & 0.008 	 & 0.017 	 & 0.010 	 &0.002 	 & 0.007 	 & 0.009 	 & 0.005 	 & \multirow{2}{*}{$+$0.046} 	 & 0.017 	 & \multirow{2}{*}{$-$0.032} 	 & 0.094 \\ 
 	 & 	 & 	 & 	 & 	 & 0.041 	 & 0.007 	 & 0.010 	 & 0.011 	 & 0.013 	 & 0.012 	 & 0.010 	 & 0.010 	 & $-$0.007 	 & 	 & 0.019 	 & 	 & 0.053 \\ 
\\
\multirow{2}{*}{Lee $\alpha$-poor} 	 & \multirow{2}{*}{345} 	 & \multirow{2}{*}{3528} 	 & \multirow{2}{*}{$+$0.027} 	 & \multirow{2}{*}{$-$0.126} 	 & 0.082 	 & 0.004 	 & 0.005 	 & 0.007 	 & 0.006 	 &0.003 	 & 0.002 	 & 0.004 	 & 0.070 	 & \multirow{2}{*}{$-$0.129} 	 & 0.040 	 & \multirow{2}{*}{$+$0.003} 	 & 0.116 \\ 
 	 & 	 & 	 & 	 & 	 & 0.041 	 & 0.001 	 & 0.010 	 & 0.013 	 & 0.007 	 & 0.005 	 & 0.007 	 & 0.003 	 & 0.004 	 & 	 & 0.034 	 & 	 & 0.057 \\ 
\\
\multirow{2}{*}{$\alpha$-bin 1} 	 & \multirow{2}{*}{137} 	 & \multirow{2}{*}{860} 	 & \multirow{2}{*}{$-$0.008} 	 & \multirow{2}{*}{$-$0.001} 	 & 0.076 	 & 0.016 	 & 0.019 	 & 0.010 	 & 0.009 	 &0.009 	 & 0.004 	 & 0.007 	 & 0.066 	 & \multirow{2}{*}{$-$0.190} 	 & 0.037 	 & \multirow{2}{*}{$+$0.189} 	 & 0.112 \\ 
 	 & 	 & 	 & 	 & 	 & 0.077 	 & 0.012 	 & 0.021 	 & 0.021 	 & 0.013 	 & 0.000 	 & 0.015 	 & 0.011 	 & $-$0.048 	 & 	 & 0.043 	 & 	 & 0.108 \\ 
\\
\multirow{2}{*}{$\alpha$-bin 2} 	 & \multirow{2}{*}{287} 	 & \multirow{2}{*}{3415} 	 & \multirow{2}{*}{$-$0.043} 	 & \multirow{2}{*}{$-$0.024} 	 & 0.044 	 & 0.010 	 & 0.014 	 & 0.020 	 & 0.007 	 &0.005 	 & 0.011 	 & 0.011 	 & 0.027 	 & \multirow{2}{*}{$+$0.020} 	 & 0.054 	 & \multirow{2}{*}{$-$0.044} 	 & 0.082 \\ 
 	 & 	 & 	 & 	 & 	 & 0.078 	 & 0.006 	 & 0.015 	 & 0.019 	 & 0.012 	 & 0.003 	 & 0.007 	 & 0.005 	 & $-$0.067 	 & 	 & 0.090 	 & 	 & 0.139 \\ 
\\
\multirow{2}{*}{$\alpha$-bin 3} 	 & \multirow{2}{*}{381} 	 & \multirow{2}{*}{3546} 	 & \multirow{2}{*}{$-$0.042} 	 & \multirow{2}{*}{$-$0.089} 	 & 0.040 	 & 0.008 	 & 0.025 	 & 0.033 	 & 0.012 	 &0.005 	 & 0.018 	 & 0.011 	 & 0.013 	 & \multirow{2}{*}{$+$0.080} 	 & 0.053 	 & \multirow{2}{*}{$-$0.169} 	 & 0.084 \\ 
 	 & 	 & 	 & 	 & 	 & 0.024 	 & 0.007 	 & 0.021 	 & 0.016 	 & 0.011 	 & 0.008 	 & 0.004 	 & 0.008 	 & $-$0.013 	 & 	 & 0.054 	 & 	 & 0.069 \\ 
\\
\multirow{2}{*}{$\alpha$-bin 4} 	 & \multirow{2}{*}{398} 	 & \multirow{2}{*}{3956} 	 & \multirow{2}{*}{$-$0.003} 	 & \multirow{2}{*}{$-$0.003} 	 & 0.089 	 & 0.018 	 & 0.000 	 & 0.000 	 & 0.012 	 &0.007 	 & 0.003 	 & 0.023 	 & 0.002 	 & \multirow{2}{*}{$-$0.044} 	 & 0.055 	 & \multirow{2}{*}{$+$0.040} 	 & 0.110 \\ 
 	 & 	 & 	 & 	 & 	 & 0.041 	 & 0.011 	 & 0.092 	 & 0.052 	 & 0.051 	 & 0.007 	 & 0.034 	 & 0.019 	 & $-$0.009 	 & 	 & 0.070 	 & 	 & 0.149 \\ 
\\
\multirow{2}{*}{$\alpha$-bin 5} 	 & \multirow{2}{*}{216} 	 & \multirow{2}{*}{1092} 	 & \multirow{2}{*}{$+$0.063} 	 & \multirow{2}{*}{$+$0.044} 	 & 0.099 	 & 0.015 	 & 0.057 	 & 0.049 	 & 0.008 	 &0.012 	 & 0.016 	 & 0.032 	 & 0.013 	 & \multirow{2}{*}{0.121} 	 & 0.037 	 & \multirow{2}{*}{$-$0.078} 	 & 0.137 \\ 
 	 & 	 & 	 & 	 & 	 & 0.118 	 & 0.030 	 & 0.010 	 & 0.013 	 & 0.025 	 & 0.005 	 & 0.020 	 & 0.016 	 & 0.045 	 & 	 & 0.038 	 & 	 & 0.141 \\ 
\enddata 
\tablecomments{
Same as Table\,\ref{tab:correlated_errors} except here all subsamples are limited to distances between 1.447 and 1.614 kpc, i.e., are volume-complete over the distance range of the Full Sample. Note that the number of stars decrease significantly for most samples, increasing the overall uncertainties on the slope.}
\end{deluxetable}
\end{landscape}

\section{Vertical Metallicity Gradient of Different Subsamples}
\label{sec: vmg_results} 

\subsection{Disk as a Whole} 
\label{sec:whole_disk}

\begin{figure}[htbp]
\begin{center}
\includegraphics[width=\textwidth]{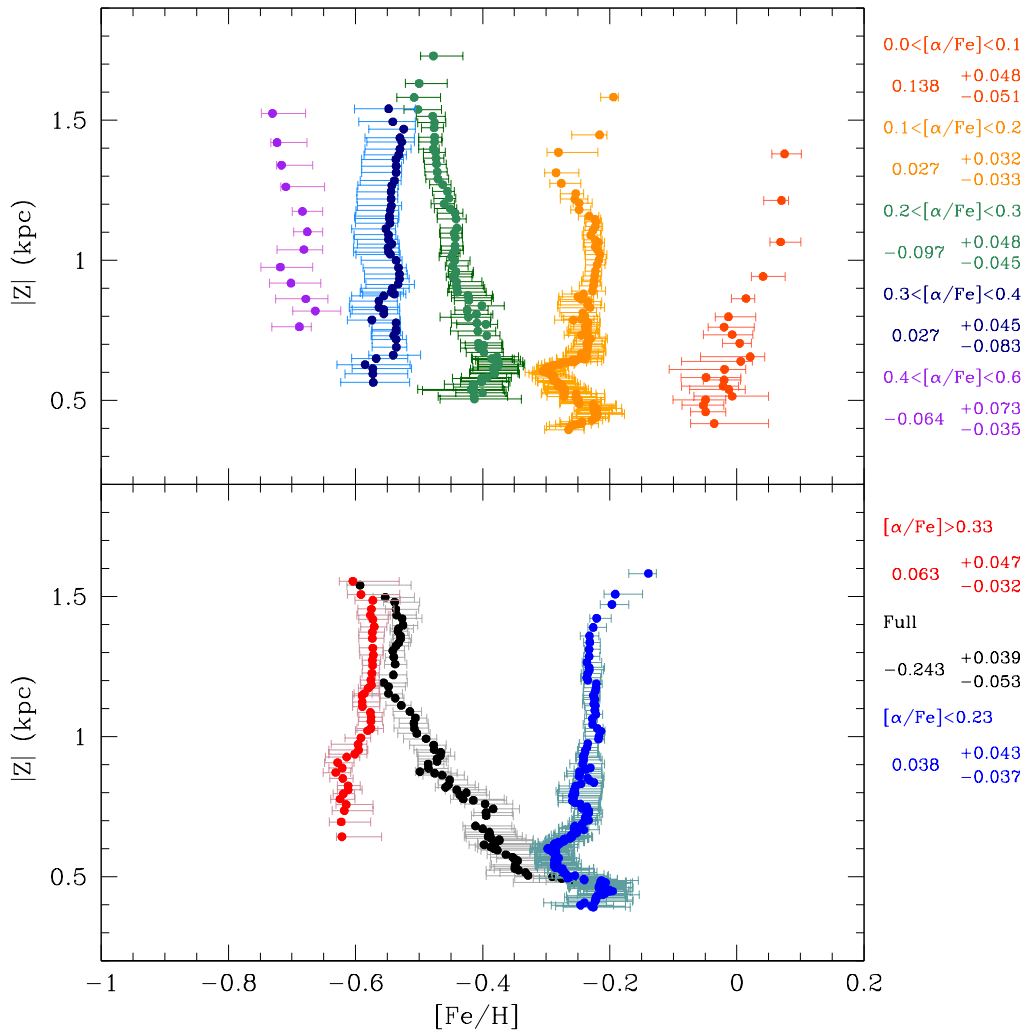}
\caption{$Top:$ The vertical metallicity gradients of the different [$\alpha$/Fe] subsamples. The error bars on each point reflect the uncertainties described in \S\,\ref{sec:grad_uncertainties}, and the slopes have been adjusted for the correlated errors in [Fe/H] and distance. The different colors are for the different $\alpha$-bin subsamples, and the legend and gradients are listed on the right-hand side. Each of these subsamples is volume-complete over the distance ranges listed in Table\,\ref{tab:dist_limits}. $Bottom:$ A similar figure except for different [$\alpha$/Fe] ranges. The black points are the full SEGUE G-dwarf sample, the red are the $\alpha$-rich subsample, and the blue are for the $\alpha$-poor subsample. 
}
\label{fig:vgrad_slopes}
\end{center}
\end{figure}

Using our boxcar-smoothing technique and various systematic corrections, we have measured a vertical metallicity gradient of $-$0.243$^{+0.039}_{-0.053}$ dex kpc$^{-1}$ for the volume-complete Full Sample of SEGUE G dwarfs from 0.27 to 1.62 kpc in $|Z|$ (bottom panel of Figure\,\ref{fig:vgrad_slopes}, Table\,\ref{tab:correlated_errors}). This volume-complete sample contains stars with metallicity in the range $-$1.000 $<$ [Fe/H] $< -$0.035 and $\alpha$-enhancement 0.00 $<$ [$\alpha$/Fe] $<$ +0.55. At low heights, the stars are metal-rich ([Fe/H]$ \approx -$0.3); as $|Z|$ increases, the sample becomes increasingly metal-poor, levelling off around an [Fe/H] = $-$0.55. 

There are some ``wiggles" in [Fe/H] around $|Z|$ of 0.6 and 0.8 kpc. These ``wiggles" are well within our error bars and are a manifestation of large target-selection weights for a few stars at this height; although larger than average, these weights are still within our 2$\sigma$ limit. This ``bend" is not seen in the raw sample (see Figure\,\ref{fig:triple_whammy}), and when we use a more stringent cut on the target-selection weights, the feature disappears. We do not believe that this behavior shows any peculiar structure at this height but rather arises from our small sample size close to the plane of the Galaxy. As indicated in Figure\,\ref{fig:vgrad_slopes} and discussed in Appendix\,\ref{appendix}, the uncertainties at these low heights are larger than at higher $|Z|$. 

\subsection{$\alpha$-subsamples} 

We have also divided the SEGUE G-dwarf sample into bins of [$\alpha$/Fe], each of which probes a different volume-complete distance range (Table\,\ref{tab:dist_limits}). Examining these subsamples provides a more detailed picture of the disk and does not pre-suppose a thin/thick disk structure. The top panel of Figure\,\ref{fig:vgrad_slopes} shows the measured vertical metallicity gradient for our volume-complete [$\alpha$/Fe] subsamples after our corrections for target-selection biases and correlated errors have been applied. These values are also listed in Table\,\ref{tab:correlated_errors}. 

Our lowest $\alpha$-bin subsample, with 0.0$ < $[$\alpha$/Fe]$ < $+0.1, has a positive vertical metallicity gradient, $+$0.138$^{+0.048}_{-0.051}$ dex kpc$^{-1}$. While the weighted vertical metallicity gradient itself is quite flat, with a slope of 0.009 dex kpc$^{-1}$, factoring in the correlated errors shifts the slope in the positive direction. As noted in \S\,\ref{sec:corr_errors}, even small changes in [Fe/H] at the metal-rich end will result in large changes in distance. Thus, correlated errors have a larger affect on the gradient of this sample than the other $\alpha$-bins. Note also that this subsample has the smallest number of stars and is in a region of the Galaxy that is not as well-sampled by the SEGUE survey. Thus, the magnitude of this gradient is uncertain.
 
As the sample becomes increasingly $\alpha$-rich, the vertical metallicity gradient changes. For $\alpha$-bin 2, with +0.1$ < $[$\alpha$/Fe]$ < $+0.2, the slope is $+$0.027$^{+0.032}_{-0.033}$ dex kpc$^{-1}$. $\alpha$-bin 4 shows a similar value, with a slope of $+$0.027$^{+0.045}_{-0.083}$ dex kpc$^{-1}$. In contrast, for $\alpha$-bins 3 and 5, we estimate slopes of $-$0.097$^{+0.048}_{-0.045}$ and $-$0.064$^{+0.073}_{-0.035}$ dex kpc$^{-1}$, respectively. Although the sign and extent of the vertical metallicity gradients for the $\alpha$-subsamples varies, each shows minimal change in median [Fe/H] with respect to height above the Galactic plane. With the exception of $\alpha$-bin 1, the measured vertical gradients are generally consistent with one another within the expected uncertainties; they are also consistent within 2$\sigma$ with a flat metallicity gradient. Thus, while we see a strong vertical metallicity gradient over the disk as a whole, individual $\alpha$-populations show little variation in the median [Fe/H] with height above the Galactic plane.

\begin{figure}[htbp]
\begin{center}
\includegraphics[width=\textwidth]{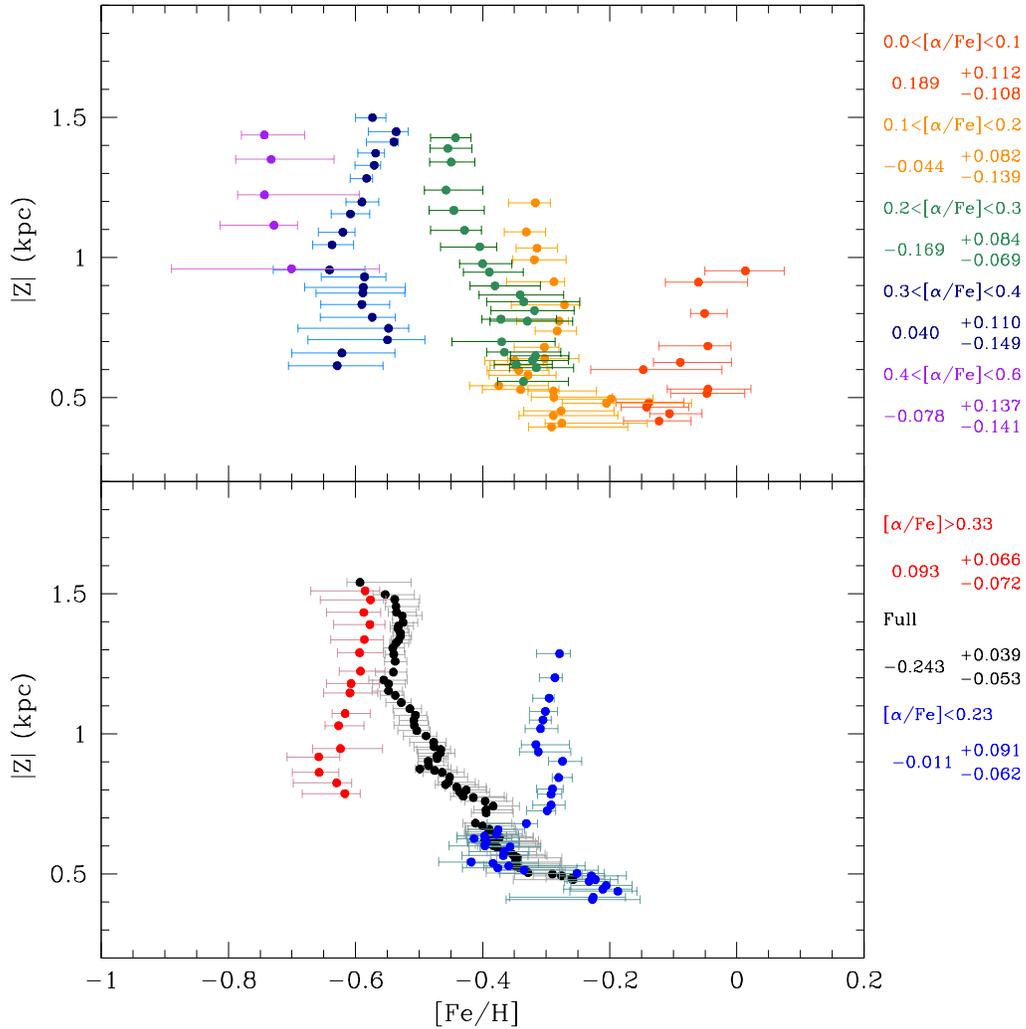}
\caption{Same as Figure\,\ref{fig:vgrad_slopes} except that, rather than covering the distance ranges listed in Table\,\ref{tab:dist_limits}, here all subsamples are over the volume-complete distance range of the Full Sample, from 1.447 to 1.614 kpc. The number of stars in each of these subsamples is decreased by about a third of those represented in Figure\,\ref{fig:vgrad_slopes}, producing much larger uncertainties on the estimated gradients. 
}
\label{fig:vgrad_slopes2}
\end{center}
\end{figure}

We also examine the vertical metallicity gradients of the $\alpha$-bins over the volume-complete distance range for the Full Sample, from 1.447 to 1.614 kpc, which covers from 0.28 to 1.58 kpc in $|Z|$ (Figure\,\ref{fig:vgrad_slopes2}). These two sets of vertical metallicity gradients provide slightly different information. The gradients discussed above, with each subsample covering the distance range specified by Table\,\ref{tab:dist_limits}, allow us to examine each individual [$\alpha$/Fe] population over as much of the disk as possible with SEGUE. Namely, we can inspect the variations in [Fe/H] over the largest possible disk volume. This approach also includes the largest possible number of stars for each subsample. In contrast, when we limit the subsamples to distances between 1.447 to 1.614 kpc, we can directly compare the different gradients with one another (Table\,\ref{tab:correlated_errors_overlap}). The bias in metallicity with respect to volume coverage is minimised (\S\,\ref{sec:vol_complete}), and we can examine the interplay of different chemical populations over a limited disk volume. With the exception of $\alpha$-bin 1, the gradients of each subsample over the two distance cuts are consistent within 1$\sigma$ and a slope of 0.0 within 2$\sigma$. As with our other distance cuts, the gradient of each $\alpha$-bin subsample is much flatter than that of the disk as a whole. Of the different bins, the sample with [$\alpha$/Fe] between +0.2 and +0.3 exhibits the strongest gradient over the Full Sample distance range; this population covers the bimodal break in [$\alpha$/Fe], which may explain some of this structure (see Figure\,\ref{fig:afefeh_scatter}). 

Examining the vertical metallicity gradient for each subsample over the volume-complete distance range of the Full Sample reveals that different $\alpha$ populations are present at different heights (Figure\,\ref{fig:vgrad_slopes2}). For example, while there are a negligible number of stars with [$\alpha$/Fe]$ < +$0.1 above $|Z| =$ 1 kpc, there are also few stars with +0.4$ < $[$\alpha$/Fe]$ < $ +0.6 below this height. Each [$\alpha$/Fe] population probes a different range of $|Z|$ heights. We discuss the repercussions of this result in \S\,\ref{sec:scale_heights}. 

\subsection{``Thin" and ``Thick" disk comparison} 

The early picture of the disk of the Milky Way consists of an $\alpha$-rich thick disk and an $\alpha$-poor thin disk (e.g., \citealt{fuhrmann98, fuhrmann11}). For the sake of comparison with previous analyses, we divide our sample of SEGUE G-dwarfs into two larger bins of [$\alpha$/Fe] to examine the vertical structure of these two proposed components. If there are two separable components, with distinct formation and evolution processes, they may exhibit different vertical metallicity structure.

First, we divide the sample according to the chemical separation defined by \citet{lee11b}. Using our boxcar smoothing, we measure the corrected vertical metallicity gradients for these subsamples over a volume-complete distance range listed in Table\,\ref{tab:dist_limits}. As with the smaller [$\alpha$/Fe] bins, the gradients flatten significantly compared to that over the disk as a whole. The $\alpha$-rich stars have a slope of $-$0.020$^{+0.035}_{-0.033}$ dex kpc$^{-1}$. The $\alpha$-poor sample has a slope of 0.087$^{+0.074}_{-0.059}$ dex kpc$^{-1}$ (Table\,\ref{tab:correlated_errors}). Both of these subsamples exhibit little change in median [Fe/H] with increasing distance from the Galactic plane. 

As noted earlier, the \citet{lee11b} cuts were based on the [$\alpha$/Fe] vs [Fe/H] distribution of their raw G-dwarf sample. However, these distributions change when we account for the SEGUE target-selection biases (Figure\,\ref{fig:afefeh_scatter}, \S\,\ref{sec:chem_subsamples}). We have defined our own $\alpha$-rich and $\alpha$-poor populations based on the unbiased distribution in chemical space and determined vertical metallicity gradients over a volume-complete distance range. Stars with [$\alpha$/Fe]$ \geq $+0.33, associated with the thick disk, have a vertical metallicity gradient of $+$0.063$^{+0.047}_{-0.032}$ dex kpc$^{-1}$ (Figure\,\ref{fig:vgrad_slopes}). This value is larger than that obtained when using the \citet{lee11b} criteria. The vertical metallicity gradient of the $\alpha$-poor subsample, with [$\alpha$/Fe]$ \leq $+0.23, is $+$0.038$^{+0.043}_{-0.037}$; this measurement is in agreement with the \citet{lee11b} subsample value within the uncertainties. The measured vertical metallicity gradient changes slightly as we vary our cuts in chemistry. However, they show consistent overall behavior. Namely, they are all significantly flatter than the gradient over the disk as a whole and indicates that there is little change in the typical [Fe/H] with increasing $|Z|$ over a small range in [$\alpha$/Fe].

Figure\,\ref{fig:vgrad_slopes} shows that the $\alpha$-rich subsample shows a median [Fe/H] of $\approx-$0.6 from $|Z|$ of 0.5 to 1.6 kpc. The $\alpha$-poor subsample is more complex, with a large amount of structure below $|Z|$ of 0.75 kpc. While some of this complexity is due to anomalous weights (\S\,\ref{sec:whole_disk}), the variation is also due to our small sample size of stars at these low heights. With a limited number of targets, uncertainties from bootstrapping increase (Appendix\,\ref{sec:bootstrap}). Between 0.75 and 1.5 kpc in $|Z|$, where this population is better sampled by SEGUE, the $\alpha$-poor stars show little variation in the median [Fe/H], around $-$0.225. 

When we limit the distance range of our subsamples to 1.447 to 1.614 kpc, we measure gradients that are consistent within 1$\sigma$ with those using the distance limits in Table\,\ref{tab:dist_limits} (Table\,\ref{tab:correlated_errors_overlap}, Figure\,\ref{fig:vgrad_slopes2}). This change significantly decreases the sample size, resulting in larger uncertainties, particularly at low $|Z|$ for the $\alpha$-poor sample, a population which is not as well-sampled by SEGUE. Both the $\alpha$-rich and $\alpha$-poor subsamples continue to display little change in their median [Fe/H] with respect to height. As seen in the $\alpha$-bin analysis, different $\alpha$-samples dominate at different heights above the plane. There are few $\alpha$-rich stars below $|Z| =$ 0.7 kpc and few $\alpha$-poor above $|Z| =$ 1.3 kpc. The vertical metallicity gradient of the full sample is in good agreement with the $\alpha$-poor sample at low $|Z|$, where these stars dominate the population, and shifts to match the $\alpha$-rich sample at high $|Z|$. We discuss this topic further in \S\,\ref{sec:scale_heights}.

\section{Discussion} 
\label{sec: discussion}

\subsection{Comparison with Previous Observational Work} 
\label{sec: vmg_comparison} 

SEGUE provides a large sample with accurate spectroscopic stellar parameters. We can adjust the sample such that it is unbiased in chemistry, reflecting the underlying stellar populations over an extensive volume of the disk. Here we compare our results to values from the recent literature. Our results are consistent with many of these analyses, such as \citet{ivezic08}. Others show more discrepant values for the vertical metallicity gradient, whether due to limited sample size and/or spatial coverage (e.g., \citealt{kordopatis11}) or issues related to target-selection biases (e.g., \citealt{carrell12}). Of particular interest are comparisons with recent results from APOGEE \citep{hayden13} and RAVE \citep{boeche14}. 
  
\subsubsection{\citet{ivezic08}} 

Using SDSS photometry, \citet{ivezic08} determined the vertical metallicity gradients of the disk and halo. Their unbiased sample contained over 2 million F and G stars over a large volume. However, it relied on photometric metallicity indicators, which are more uncertain than spectroscopic values. In Figure\,\ref{fig:vbcpu3_noacut_lit}, we compare our measured vertical metallicity gradient for the full sample to the result of \citet{ivezic08}. There is good general agreement with their estimate. However, our gradient flattens above 1.2 kpc, whereas theirs shows this behavior only above heights of around 3 kpc. They also have a typical [Fe/H] = $-$0.75 at large heights, whereas our sample flattens to a median value around [Fe/H] = $-$0.55. This difference is likely a consequence of different [Fe/H] criteria for the two samples. While their sample of ``disk" stars extends as low as [Fe/H] = $-$1.5, we limit our sample to [Fe/H]$ \geq -$1.0, to avoid halo stars. Thus, it is not surprising that our median values at large $|Z|$ are higher than theirs. 

At low heights, our G-dwarf sample is more metal-rich than that of \citet{ivezic08}; they find a typical [Fe/H] = $-$0.4 at $|Z|$ of 0.5 kpc, whereas we measure [Fe/H] of $\approx-$0.30. Local stellar samples, such as the cool dwarfs in RAVE which probe distances from 50 to 250 pc, exhibit a metallicity around [Fe/H] = $-$0.2 at low latitudes \citep{siebert11}.  Our gradient predicts a value of approximately [Fe/H] = $-$0.21 at a height of 150 pc, in agreement with the RAVE sample, whereas the disk gradient from \citet{ivezic08} has an [Fe/H] of around $-$0.30. This result suggests that there may be systematic offsets between the SSPP metallicities and those determined using photometric indicators for metal-rich stars; similar conclusions were reached by \citet{lee11b}. 

\subsubsection{\citet{kordopatis11}} 

\citet{kordopatis11} use a sample of $\approx$700 stars along a single line of sight to examine the vertical metallicity gradient from 1$\leq|Z|\leq$4 kpc. These stars have low-resolution (R$\approx$6500) spectra and are selected based on their V magnitude and inclusion in the \citet{ojha96} sample, which provides proper motion information. Note that the \citet{ojha96} sample uses a (B-V) color criteria to identify F and G stars for analysis. 

By limiting their $|Z|$ range to above 1 kpc, \citet{kordopatis11} seek to isolate the thick disk population, which has a scale height of around 900 pc \citep{juric08}. Although the efficacy of this approach is unclear, we have analysed a subsample of the Full Sample with $|Z|>$1 kpc. Accounting for target-selection biases and correlated errors, the slope is $-$0.233$^{+0.209}_{-0.0149}$ dex kpc$^{-1}$ (Table\,\ref{tab:correlated_errors}). Figure\,\ref{fig:vbcpu3_noacut_zs1_lit} compares their measured vertical metallicity gradient to ours for $|Z|\geq$1 kpc. Their sample has a slope of $-$0.14$\pm$0.05 dex kpc$^{-1}$ between 1 and 4 kpc in $|Z|$. Due to the large uncertainties on our vertical metallicity gradient for this subsample, the two gradients are consistent with one another within 1$\sigma$. However, Figure\,\ref{fig:vbcpu3_noacut_zs1_lit} indicates that their sample is more metal-rich than ours at all heights. 

Although they do not include stars below $|Z|$ of 1 kpc in their gradient calculation, there are some in their sample; using their low $|Z|$ points, we calculate the vertical metallicity gradient for the disk as a whole for their entire sample. Applying a linear least-squares analysis, this vertical metallicity gradient is $-$0.124 dex kpc$^{-1}$ for their full sample between 0 and 4 kpc in $|Z|$ (Figure\,\ref{fig:vbcpu3_noacut_lit}). This slope is less than that of our Full Sample ($-$0.243$^{+0.039}_{-0.053}$ dex kpc$^{-1}$), and, as with the slopes above $|Z|$ of 1 kpc, their points are more metal-rich than ours. 

We suspect that the variation in the vertical metallicity gradient between their analysis and ours stems from sample differences, i.e., they are more limited in size and spatial coverage, while our sample is larger and multi-directional. Using simulated lines of sight from the model of \citet{schonrich09a, schonrich09b}, we compare the vertical metallicity gradient of the model as a whole to that of subsamples comparable to that of \citet{kordopatis11}. We manufacture 10 bootstrap with replacement versions of the \citet{schonrich09a, schonrich09b} model of the different SEGUE lines of sight. First, we determine the gradients for this simulation using our boxcar-smoothing technique and a least-squares analysis; the uncertainty is from the variation in the gradients over the 10 iterations. This is our ``control" gradient. We repeat this analysis, limiting the sample used to determine the gradient to 700 randomly selected stars, comparable to the sample size of \citet{kordopatis11}. Finally, we limit the sample in both number $and$ lines of sight. SEGUE does not contain the particular lines of sight in \citet{kordopatis11}, so we isolate stars in the same area of the sky. Applying the limits associated with our Full and $|Z|>$1 kpc samples, we find that limitations in coverage and number will affect the measured vertical metallicity gradients in a non-negligible way, likely creating the differences between our values and those of \citet{kordopatis11}. In addition, there may be some selection biases in the sample due to the color criteria used by \citet{ojha96}.

\subsubsection{\citet{katz11}} 

\begin{figure}[htbp]
\begin{center}
\includegraphics[width=0.8\textwidth]{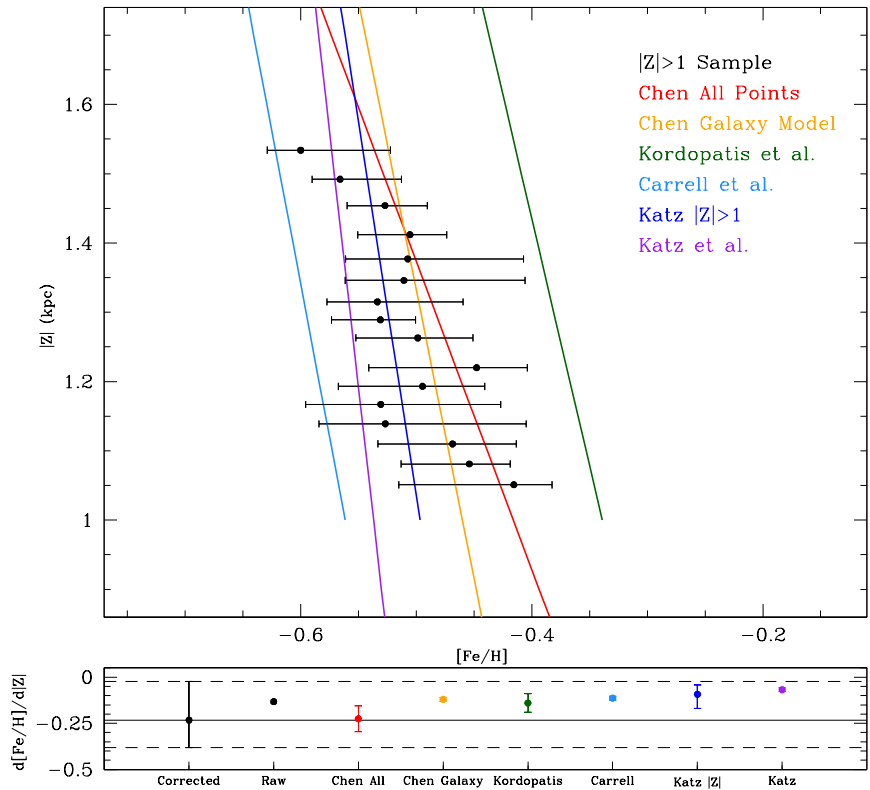}
\caption{{$Top:$ Our measured vertical metallicity gradient of $-$0.233$^{+0.209}_{-0.149}$ dex kpc$^{-1}$  for the G-dwarf sample with $|Z|\geq$1 kpc is shown in black. The gradient has been corrected for correlated errors; the uncertainties on each point reflect errors described in \S\,\ref{sec:grad_uncertainties}. The measured vertical metallicity gradients from previous analyses are given in different colors, as listed in the top right of the figure. 
$Bottom:$ A direct comparison of the different slopes from each of these analyses. The color scheme is the same as in the panel above. The solid line presents our measured slope for $|Z| > $1 kpc; the dashed show the uncertainties in this value. Due to low sample size, our uncertainties are large, and all literature values agree within 1$\sigma$. We also display the measured vertical metallicity gradient for our raw sample, before correcting for target-selection biases and correlated errors. The \citet{carrell12} value ($-$0.113$\pm$0.010 dex kpc$^{-1}$) is in agreement within 2$\sigma$ with that of our raw sample ($-$0.133 dex kpc$^{-1}$). } 
}
\label{fig:vbcpu3_noacut_zs1_lit}
\end{center}
\end{figure}

\citet{katz11} have low-resolution spectra of 400 stars in two lines of sight at a high and intermediate Galactic latitude. They use a ($B-V$) color criteria to isolate sub-giant and giant stars and design their fields to focus on stars with magnitudes between 15 and 16. Their metallicity distribution functions with respect to height show bimodality, which they exploit to isolate the thick disk from the thin disk and halo.  They find a vertical metallicity gradient of $-$0.068$\pm$0.009 dex kpc$^{-1}$ for their ``thick disk" subsample up to a $|Z|$ of 3 kpc (Figure\,\ref{fig:vbcpu3_noacut_zs1_lit}). This value is smaller than our $|Z|>$1 kpc gradient and shallower than the slope of our subsample with [$\alpha$/Fe]$ \geq $+0.33, chemically associated with the thick disk. In addition, Figure\,\ref{fig:vbcpu3_noacut_zs1_lit} indicates that their sample is slightly more metal poor than ours. 

For a more direct comparison of the two analyses, we run our boxcar-smoothing methodology on their sample of stars, which is publicly available. For stars with $|Z|\geq$1 kpc, we measure a vertical metallicity gradient of $-$0.093$^{+0.051}_{-0.077}$ dex kpc$^{-1}$, smaller than that determined for the SEGUE sample\footnote[4]{We do not know all of the details of the \citet{katz11} sample and are thus limited to a more basic error analysis. The reported errors on the slope are produced by a bootstrap with replacement analysis over 100 iterations.} (Figure\,\ref{fig:vbcpu3_noacut_zs1_lit}, Katz $|Z|>$1).  Although more metal-poor and with a smaller gradient value, this sample is consistent within 1$\sigma$ with our gradient above $|Z| =$ 1 kpc. 

Applying our boxcar smoothing technique on their sample as a whole produces a gradient of $-$0.206$^{+0.092}_{-0.068}$ dex kpc$^{-1}$ (Figure\,\ref{fig:vbcpu3_noacut_lit}). While their slope is in agreement with our values for the Full Sample, their sample is more metal rich. This contrasts with their reported thick-disk gradient, which is more metal poor than our own (Figure\,\ref{fig:vbcpu3_noacut_zs1_lit}). 

Our reanalysis of the \citet{katz11} sample is roughly consistent with our SEGUE sample in slope. However, their published gradient shows larger discrepancies with our own sample (Katz et al. in Fig.\,\ref{fig:vbcpu3_noacut_zs1_lit}). As with \citet{kordopatis11}, we expect that limitations in number and line-of-sight coverage causes some of these discrepancies. Also, it is unclear as to whether or not there are selection biases occurring in this sample and/or an offset in their metallicity determinations from those of the SSPP. 

\subsubsection{\citet{chen11}} 

To investigate the vertical metallicity gradient of the thick disk, \citet{chen11} extract a sample of 1728 red horizontal-branch (RHB) stars from SDSS DR8. These stars have a complex selection function, based upon a color-metallicity relation and magnitude cuts; they are also trimmed to have $|Z| < $3 kpc. They use two different methodologies to isolate the thick disk; geometric decomposition and removing stars from other components using Galaxy models. 

When \citet{chen11} limit their sample to stars between 1 and 3 kpc above the Galactic plane, they measure a vertical metallicity gradient of $-$0.225$\pm$0.07 dex kpc$^{-1}$ (Figure\,\ref{fig:vbcpu3_noacut_zs1_lit}, Chen All Points), showing good agreement with our results above $|Z| =$ 1 kpc. 

To isolate the thick disk an alternate way, they model the distribution of thin-disk and halo stars that meet their RHB selection criteria with the Besancon Galaxy Model \citep{robin03}. Scaling these distributions to their RHB sample size, they remove these two components to isolate the thick-disk stars with 0.5$<|Z|<$3.0 kpc.  This subsample exhibits a smaller gradient, $-$0.12$\pm$0.01 dex kpc$^{-1}$ (Figure\,\ref{fig:vbcpu3_noacut_zs1_lit}, Chen Galaxy Model), than that from geometric decomposition. They suspect that by removing the thin-disk and halo components, they may have artificially removed the more extreme metallicities from the thick disk, resulting in a smaller measured gradient. As expected, this results in a smaller gradient than our estimates for stars with $|Z| > $1 kpc. 

\subsubsection{\citet{carrell12}} 

\citet{carrell12} measured the vertical metallicity gradient of 43,417 SEGUE F, G, and K dwarfs. To isolate the thick disk, \citet{carrell12} selected stars with 1$<|Z|<$3 kpc, measuring a gradient of $-$0.113$\pm$0.010 dex kpc$^{-1}$ for targets with 7$<R_{GC}<$10.5 kpc. Using a different distance determination method, they find a slightly larger gradient of $-$0.125$\pm$0.008 dex kpc$^{-1}$. Both of these values are smaller than our measured value for the vertical metallicity gradient over the span of the disk. These differences in the measured gradients stem from uncorrected selection biases in the SEGUE sample.

As explained in \S\,\ref{sec:selection_biases}, while we select all stars that fulfil the appropriate color and magnitude criteria, including those targeted under different SEGUE categories, \citet{carrell12} isolate stars that meet the SEGUE selection criteria of {\it only} F, G, or K dwarfs. In our analysis, we account for the biases that stem from overlapping target-selection categories. In contrast, they need to account for the biases that will stem from {\it avoiding} these other categories, which will bias their sample against metal-poor stars. Furthermore, their sample will also suffer from the target-selection biases related to variations in the stellar density in fields of different latitude, which will lead to a dramatic over-representation of metal-poor stars. Finally, they do not take into account the effect of correlated errors on the metallicity gradient. As shown in Figure\,\ref{fig:triple_whammy}, these adjustments dramatically affect the measured vertical metallicity gradients. Above $|Z|$ of 1 kpc, the \citet{carrell12} values are in good agreement with our unweighted vertical metallicity gradient, which has a slope of $-$0.133 dex kpc$^{-1}$ (Figure\,\ref{fig:vbcpu3_noacut_zs1_lit}, Table\,\ref{tab:correlated_errors}). By not accounting for the various SEGUE target-selection biases, they estimate gradients much smaller than our values.

The \citet{carrell12} gradient is also offset in [Fe/H] from our weighted slope. With a value of [Fe/H] = $-$0.45, the intercept for their gradient is close to that of our unweighted gradient, which is [Fe/H] $\approx$ $-$0.35 (Figure\,\ref{fig:triple_whammy}). This offset is also likely related to target-selection biases in their sample, namely the bias towards metal-poor stars from SEGUE's sampling of equal number of stars in fields regardless of their stellar density. 

\subsubsection{\citet{allendeprieto06}}

\citet{allendeprieto06} extracted a sample of $\sim$23,000 F- and G-type stars from SDSS Data Release 3 \citep{abazajian05}. Using their own analysis pipeline, distinct from the SSPP, they separate the spectral types based on atmospheric parameters and examine the metallicity structure of the disk and halo. Similar to SEGUE, this sample is biased against metal-rich stars. 

Examining the vertical metallicity gradient between $|Z|$ of 1 and 3 kpc for G-type stars with [Fe/H]$ > -$1.2, \citet{allendeprieto06} report that the vertical metallicity gradient must be less than 0.03 dex kpc$^{-1}$, much smaller than our estimated value for the disk as a whole. However, a metallicity distribution function (MDF) of their sample suggests that their stars are primarily from the thick disk. They find a median [Fe/H] = $-$0.679, more metal-poor than our measured median value for the sample as a whole. When we compare their gradient with that of our $\alpha$-rich subsample, which is a more appropriate match in [Fe/H] range, there is much better agreement between the two analyses. Their sample remains more metal-poor than ours, but this difference could stem from different selection criteria, namely their sample extends to lower [Fe/H] and does not use photometric target selection, or variation in the atmospheric parameters between their pipeline and the SSPP. It may also result from the bias against metal-rich stars, which they do not adjust for. 

\subsubsection{\citet{hayden13}} 

\begin{figure}[htbp]
\begin{center}
\includegraphics[width=0.9\textwidth]{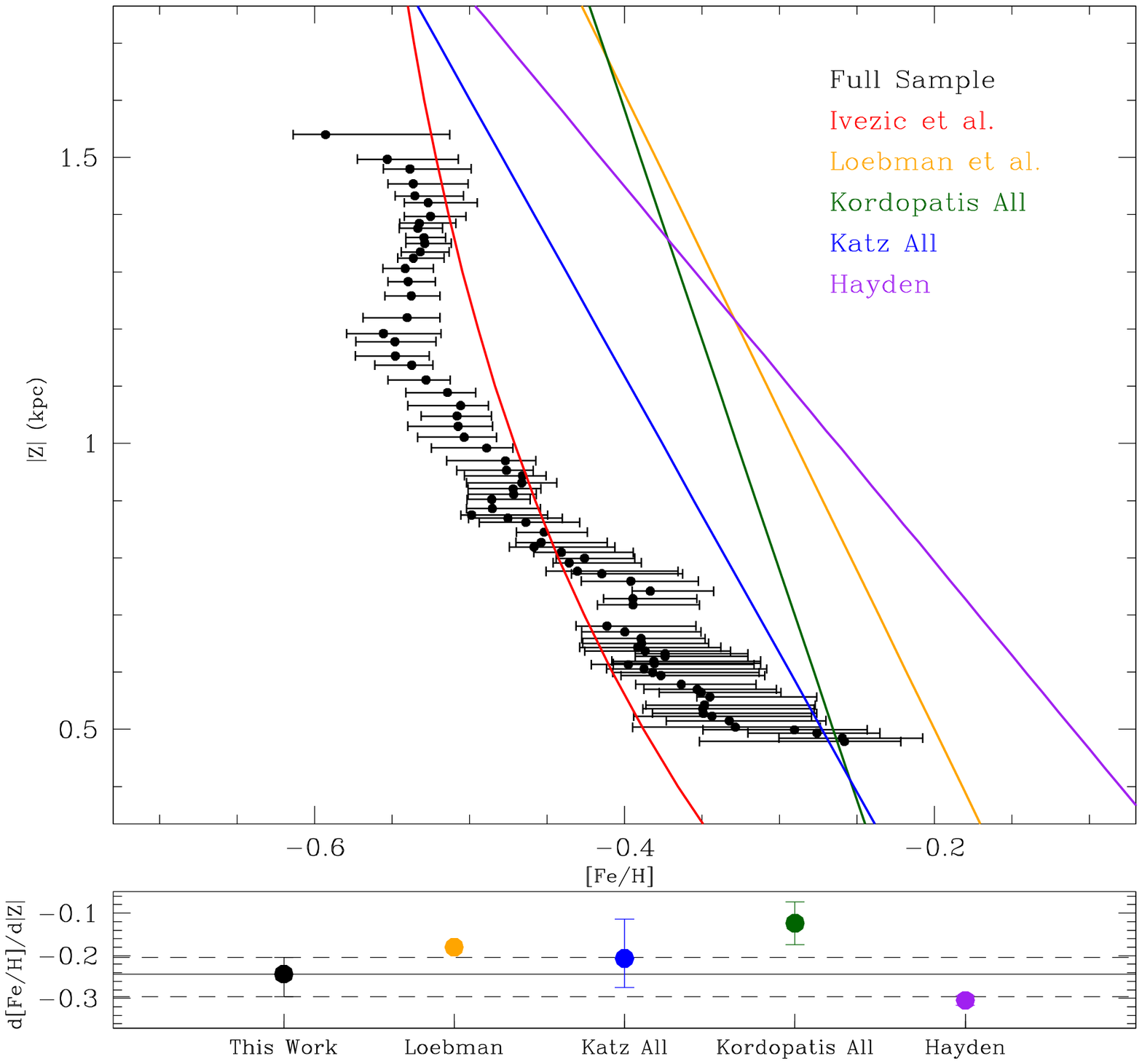}
\caption{ {$Top:$ Our measured vertical metallicity gradient of the volume-complete Full G-dwarf Sample compared with values from the literature, which are shown in different colors as listed in the top right. 
$Bottom:$ A direct comparison of the different slopes from each of these analyses. The color scheme is the same as above. The solid line shows our measured slope for the Full Sample; the dashed show the uncertainties in this value. We do not present the value from \citet{ivezic08} as they define their relationship as an exponential.  } 
}
\label{fig:vbcpu3_noacut_lit}
\end{center}
\end{figure}

\citet{hayden13} determine a vertical metallicity gradient for the first year of SDSS-III APOGEE data. Unlike our SEGUE sample, their APOGEE stars probe a wide range of Galactocentric radius, from 1 to 15 kpc. For the radial region between 7 and 9 kpc, comparable to our sample, they find a gradient of $-$0.305$\pm$0.011 dex kpc$^{-1}$ between 0 and 2 kpc in $|Z|$, larger than our estimated value (Figure\,\ref{fig:vbcpu3_noacut_lit}). Their sample also appears to be more metal-rich than ours at all heights. When they separate by $\alpha$-abundance, using criteria similar to \citet{lee11b}, they estimate gradients of $-$0.215$\pm$0.023 and $-$0.260$\pm$0.022 for the $\alpha$-poor and -rich subsamples, respectively. When they limit their sample to small bins of [$\alpha$/Fe] they measure a smaller vertical metallicity gradient of around $-$0.07 dex kpc$^{-1}$ (Hayden, personal communication). Similar to our analysis, individual [$\alpha$/Fe] populations exhibit flatter vertical metallicity gradients than the sample as a whole in their analysis.

It is currently unclear how APOGEE target selection, and the use of giants rather than dwarfs, will affect the distribution of metallicities in their sample; this may contribute to the discrepancies between the two analyses. In addition, the [$\alpha$/Fe] from SEGUE and APOGEE have yet to be compared. If there are systematic differences between the two [$\alpha$/Fe] values, that could produce a gradient for one survey that is not found in the other. Specifically, if there are stars in a given [$\alpha$/Fe] subsample whose true [$\alpha$/Fe] actually lies outside the appropriate range, they can induce a negative vertical metallicity gradient, similar to that seen over the disk as a whole. Most importantly, the two surveys probe complementary portions of the Milky Way. While SEGUE looks above the plane, much of APOGEE is focused on stars within it. Thus, APOGEE's data is best suited to the portion of the Galaxy where our SEGUE vertical metallicity gradients are the most uncertain, e.g., metal-rich stars close to the plane of the Galaxy. Conversely, SEGUE samples the high $|Z|$, $\alpha$-rich, metal-poor disk population very well, in contrast to APOGEE. Thus, that these two surveys show a similar trend for the disk as a whole and flattening for smaller bins in [$\alpha$/Fe] is quite promising. We look forward to continuing our comparison with APOGEE as they expand their sample. 

\subsubsection{\citet{boeche2014}}

Selecting a sample of 17,950 Red clump stars from RAVE, \citet{boeche2014} estimate the vertical metallicity gradient over small bins of $|Z|$. They find a slight negative gradient for heights below 0.4 kpc, which gets increasingly negative with height, reaching a value of $-$0.199$\pm$0.70 dex kpc$^{-1}$ for 1.2 $\leq |Z| \leq$ 2.0 kpc. Although they do not estimate the overall gradient, the average value for $|Z|$ between 0.4 and 2.0 kpc is approximately 0.145 dex kpc$^{-1}$, as indicated in their Figure 6. These values are consistent with our measurement of the vertical metallicity gradient of the SEGUE G-dwarf sample as a whole. It is promising that these two large surveys, which use different target-selection and stellar-analysis techniques, are consistent. 

\subsection{Comparison with Galactic Disk Models} 
\label{sec:comp_with_models} 

In models of thick disk formation by \citet{brook04, brook05} and \citet{bournaud09}, the galaxy accretes gas-rich material early on. This behavior prompts a burst of star formation which creates the thick disk either {\it in situ} or over a dynamically short time period, resulting in a vertically-uniform thick disk. Their simulations show a narrow range in [$\alpha$/Fe] associated with the simulated thick disk; most cover less than 0.15 dex, comparable to our $\alpha$-bin subsamples. They predict no vertical metallicity gradient for these simulations, in agreement with our analysis of $\alpha$-subsamples, suggesting that the disk is well-mixed chemically over $|Z|$ for a given epoch of star formation. 

If the initial disk exhibits a vertical metallicity gradient, it can be smoothed by radial migration, which will mix up stars of various chemical compositions at different $|Z|$ heights. However, if there is no initial vertical metallicity gradient, radial migration can induce one, as older stars have more time, and thus opportunity, to move outwards in $R$ and $|Z|$. The predictions from these simulations vary significantly depending on initial parameters. 

\citet{loebman11} form a thick-disk component via radial migration in an N-body simulation. Their model is not designed to recreate the Milky Way and is not calibrated with existing observational data. This radial migration will create a vertical metallicity gradient for the disk as a whole of $\approx -$0.18 dex kpc$^{-1}$, just outside our quantified uncertainties (Figure\,\ref{fig:vbcpu3_noacut_lit}). Their gradient is more metal rich than ours, which is not surprising for a model independent of observations. 

In contrast to \citet{loebman11}, the simulations of \citet{bird13} indicate that the chemical structure of the disk is largely set by {\it in situ} star formation within a rotating, collapsing gas cloud; although secular heating and radial migration are present, they have a sub-dominant effect on the chemical trends. This result suggests that our metal-poor, $\alpha$-enhanced stars formed in large scale-height populations, rather than scattering to larger $|Z|$ over time. The effect of radial migration on our sample is currently unclear due to the wide range of predicted vertical metallicity structure. As these simulations improve, they will have more detailed chemical information. The added parameter of [$\alpha$/Fe] will allow one to better distinguish between different formation models. 

\subsection{Vertical Metallicity Gradient as a Reflection of Scale Heights} 
\label{sec:scale_heights}

\begin{figure}[htbp]
\begin{center}
\includegraphics[width=\textwidth]{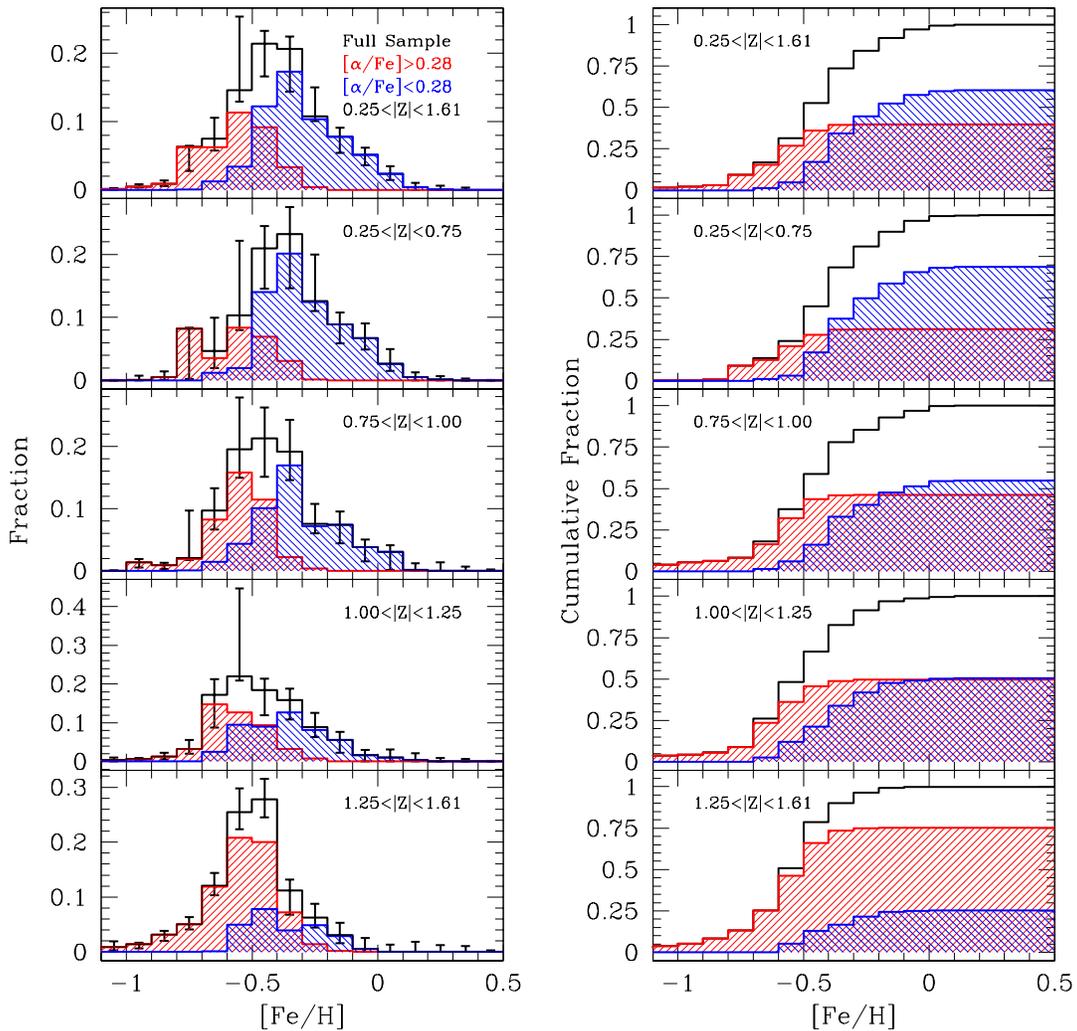} 
\caption{$Left:$ The metallicity distribution functions of three of our subsamples over a distance range where all three are volume-complete, 1.447$<$d$<$1.614 kpc. The black, with error bars, shows the Full Sample of G dwarfs. The blue and red are subsamples with [$\alpha$/Fe]$ < $+0.28 and $ > $+0.28 respectively. The top row presents the full metallicity distribution function for each subsample. Each row below shows the MDFs for different ranges of $|Z|$ in kpc, as listed in the top right corner of each panel. 
$Right:$ The cumulative distributions of the subsamples. Again, the top row is the total sample, while each of the other rows is a range of $|Z|$. 
}
\label{fig:mcpu3_hardcut_total}
\end{center}
\end{figure}

As $|Z|$ increases, we expect to detect more and more $\alpha$-enhanced stars relative to $\alpha$-poor (e.g., \citealt{bovy12a}, \citealt{bovy12b}, \citealt{schlesinger12}). Figure\,\ref{fig:mcpu3_hardcut_total} shows the metallicity distribution function of the Full, $\alpha$-rich, and $\alpha$-poor samples with respect to $|Z|$. These three samples cover a distance range where all are volume-complete (1.447$ < $d$ < $1.614 kpc); the error bars reflect uncertainties from a bootstrap (with replacement) analysis over 100 iterations (see Appendix\,\ref{sec:bootstrap}). To ensure that the number of stars in each of our $\alpha$ subsamples sum to the total number of stars, we define our $\alpha$-rich stars as having [$\alpha$/Fe]$ > $+0.28 and $\alpha$-poor as [$\alpha$/Fe]$ < $+0.28 dex for this comparison. This cut is motivated by the distribution shown in Figure\,\ref{fig:afefeh_scatter}. At low $|Z|$, the population is approximately 70\% metal-rich, $\alpha$-poor stars. As $|Z|$ increases, this percentage decreases; at $|Z|$ $\approx$ 1 kpc, the total sample is approximately 50\% $\alpha$-poor, metal-rich stars. Above 1.25 kpc, the G-dwarf sample is dominated by metal-poor, $\alpha$-enhanced stars (75\%). As the height above the Galactic plane increases, the disk transitions between [$\alpha$/Fe] populations, manifesting itself as a strong vertical metallicity gradient. 

We compare the vertical metallicity gradient of the full sample and an $\alpha$-rich and $\alpha$-poor subsample in the lower panel of Figure\,\ref{fig:vgrad_slopes2}. Each of these gradients cover the same volume-complete distance range. As seen in Figure\,\ref{fig:mcpu3_hardcut_total}, at low $|Z|$ there are few $\alpha$-rich stars; Figure\,\ref{fig:vgrad_slopes2} shows that the vertical metallicity gradient of the full sample is aligned with the $\alpha$-poor vertical metallicity gradient. At heights above $\approx$0.8 kpc, the proportion of $\alpha$-rich stars increases (as reflected in the cumulative distributions of Figure\,\ref{fig:mcpu3_hardcut_total}). The gradient of the Full Sample shifts to lower median [Fe/H]. Above 1.4 kpc, there are few stars in the $\alpha$-poor subsample, and the gradient of the Full Sample is aligned with the $\alpha$-rich. 

The top panel of Figure\,\ref{fig:vgrad_slopes2} allows further study of the variation in chemistry with height. Stars with [$\alpha$/Fe] between 0.0 and +0.1 are rarely present above 1 kpc. In contrast, stars with +0.4$ < $[$\alpha$/Fe]$ < $+0.6 have few members below this height. To further examine this point, we determine the mean $|Z|$, and $\sigma_{|Z|}$, for bins of 0.025 dex in [$\alpha$/Fe] (Figure\,\ref{fig:afezscatter}). Each bin contains a different number of targets, as indicated in the histogram in the top panel of Figure\,\ref{fig:afezscatter}. There is a clear increase in mean $|Z|$ with $\alpha$-enhancement, from around 0.5 kpc at [$\alpha$/Fe]$ = $0.0 to 1.0 kpc  at [$\alpha$/Fe]$ = $+0.5. In addition, the spread in $|Z|$ increases with $\alpha$-enhancement (bottom panel of Figure\,\ref{fig:afezscatter}). It is unclear whether $\sigma_{|Z|}$ is flat for [$\alpha$/Fe] between 0.00 and 0.15 and 0.3 and 0.5, with a transition region in the middle, or smoothly transitions over the entire [$\alpha$/Fe] range. We leave further investigation of this for a future study. 

As expected from investigation of the MDFs, the strong vertical gradient measured for the total sample reflects the transition from $\alpha$-poor, metal-rich stars at low $|Z|$ to $\alpha$-rich, metal-poor stars at high $|Z|$. Similarly, Figure\,\ref{fig:vgrad_slopes2} shows a mix of stars from [$\alpha$/Fe] of 0.1 to 0.6 above $|Z|$ of 1 kpc. Consequently, we are not surprised to find a negative vertical metallicity gradient above this height that is comparable to that of the sample as a whole. 

\subsection{Corollations with Age} 
\label{sec:age} 

The $\alpha$-abundance ratio for a given stellar population is intrinsically linked with the various timescales of chemical evolution. Due to changes in the relative contribution to the interstellar medium by Type II and Type Ia supernovae ejecta, stars formed earlier in the Galaxy have higher [$\alpha$/Fe], whereas more recently-born stars have lower [$\alpha$/Fe]. However, the purity of this relationship is unclear. The high-resolution analysis of 189 F- and G-dwarf by \citet{edvardsson93} found significant scatter in the chemistry of disk stars formed at the same time. With improved ages and chemical abundances, \citet{haywood13} find a less-scattered relationship between age and chemistry for over 1000 FGK dwarfs in the solar neighborhood, albeit with a number of significant outliers. Further complicating matters, \citet{minchev13} use a chemodynamical model to find that, while there is a clear relationship between chemistry and age for stars from a particular birth radius, radial migration can destroy all evidence of it.

If we assume that each value of [$\alpha$/Fe] is associated with a certain stellar age, our results suggest that the vertical metallicity gradient of the disk as a whole reflects changes in the chemical structure over a range of ages, while individual $\alpha$-bins reveal the behavior of a particular epoch of star formation. We find consistently small vertical metallicity gradients for our $\alpha$-subsamples, consistent with predictions of the chemodynamical model of \citet{minchev13}. First, this means that each ``age" of stars is associated with a particular median [Fe/H], regardless of its vertical location in the disk. Second, the consistency over the range of [$\alpha$/Fe] implies that each epoch of star formation creates similar vertical metallicity structure. Many models suggest that different disk-formation processes will produce different vertical chemical structure (\S\,\ref{sec:intro},\,\ref{sec:comp_with_models}); the consistency of our vertical metallicity gradient with increasing [$\alpha$/Fe] suggests that there is not a dramatic change in the dominant formation mechanisms over time.

We can also consider the links between age and typical $|Z|$ height. As described in \S\,\ref{sec:scale_heights}, stars with low [$\alpha$/Fe] are typically close to the plane of the Galaxy, whereas stars with enhanced [$\alpha$/Fe] appear at a larger mean $|Z|$, in addition to having a larger range of $|Z|$ values (Figures\,\ref{fig:afezscatter} and\,\ref{fig:vgrad_slopes2}). This indicates that stars that formed late remain at low $|Z|$, in contrast to stars that formed much earlier. These stars could be born {\it in situ} at a wider range of $|Z|$ than young stars, or they could extend their range of heights through various dynamical effects.

Recently, \citet{bird13} analysed a high-resolution hydrodynamic simulation of a Milky Way-like galaxy, which included an active merger phase at $z>$3, secular heating, and radial migration. By examining the evolution of the simulation with respect to different epochs of star formation, they determined that the disk structure formed ``upside-down." Old stellar populations formed in a vertically extended and radially compact structure; subsequent star formation shows a smooth transition to increasingly vertically compact and radially elongated spatial configurations as time progresses. \citet{stinson13} have performed a cosmological hydrodynamical simulation of a Milky Way-like galaxy. Consistent with \citet{bird13}, they find that older, more metal-poor and $\alpha$-rich stars exhibited short scale lengths and large scale heights, while younger, more metal-rich stars had long scale lengths and short scale heights. The chemodynamical disk model of \citet{minchev13} also made similar predictions. These analyses fit well with the smooth increase in scale heights with decreasing [Fe/H] and increasing [$\alpha$/Fe] found by \citet{bovy12b}. Similarly, when we assume that chemistry and age are intrinsically linked, these simulations are consistent with the variation in $|Z|$ for our different $\alpha$-subsamples and the increase in mean $|Z|$ with respect to $\alpha$-abundance (Figures\,\ref{fig:afezscatter} and\,\ref{fig:vgrad_slopes2}). 

Using high-resolution spectroscopy of a sample of $\approx$850 FGK dwarfs in the solar-neighborhood, \citet{adibekyan} reported a gap in the [$\alpha$/Fe] vs. [Fe/H] distribution (see their Figure 1). Using this stellar sample, \citet{haywood13} have proposed two distinct epochs of Milky Way star formation which give rise to the thick and thin disk, respectively. However, they are careful to note that two epochs of star formation do not necessarily give rise to two distinct disk components. Defining the ``thick disk" as consisting of stars with ages greater than 8 Gyr, they detect an age-metallicity and age-$\sigma_{W}$ relationship. These correlations mean that older, metal-poor stars will extend farther from the Galactic plane than younger, metal-rich stars, giving rise to a vertical metallicity gradient. 

Due to the low resolution of SEGUE spectra, uncertainties in chemical abundances will smear out any [$\alpha$/Fe] vs. [Fe/H] gap that may be present \citep{haywood13}; thus, we cannot separate our sample using their thin/thick disk criteria (Figure\,\ref{fig:afefeh_scatter}). However, their analysis indicates that stars with [Fe/H]$ < -$0.5 and [$\alpha$/Fe] above $\sim$0.05 should have ages larger than 8 Gyr. Thus, our $\alpha$-rich subsample, which has stars with $-$1.0$ < $[Fe/H]$ < -$0.45 and [$\alpha$/Fe]$ > $+0.33, is comparable to their ``thick disk" sample isolated by age. While they predict a negative vertical metallicity gradient for these old stars, we find a negligible change in [Fe/H] with respect to increasing $|Z|$. This result suggests that the age-metallicity and age-$\sigma_{W}$ relationships are not as clean and well-defined throughout the {\it in situ} disk as they find for their solar neighborhood sample. 

\subsection{The Thin/Thick Disk Dichotomy} 

Numerous analyses of disk chemistry find a separation in [$\alpha$/Fe] vs. [Fe/H] space (e.g., \citealt{fuhrmann98, fuhrmann11}, \citealt{lee11b}, \citealt{adibekyan}, \citealt{hayden13}, \citealt{anders13}, \citealt{ramirez13}), which is oftentimes invoked to support the separable thin/thick disk paradigm. \citet{schonrich09a}, \citet{haywood13}, and \citet{minchev13} discuss that a gap in chemical space can be a naturally occurring feature from different epochs of star formation and does not necessarily indicate two separate stellar populations. The latter picture is supported by the well-defined break in chemistry identified by \citet{haywood13} around 8 Gyr. However, the cleanliness of the age-chemistry relationship is unclear; the kinematically-defined thin and thick disk in \citet{ramirez13} show significant overlaps in [O/Fe] and [Fe/H] abundance for stars over a range of ages. While we observe chemical bimodality in our sample of G dwarfs (Figure\,\ref{fig:afefeh_scatter}), it is different than the gap observed in other data sets and at least partially stems from the line-of-sight coverage of SEGUE (\S\,\ref{sec:chem_subsamples}).

We have determined that the vertical metallicity gradient of the disk as a whole reflects the different scale heights of each [$\alpha$/Fe] population. There is no break in the behavior of the disk with respect to vertical coverage and chemistry, and the typical $|Z|$ increases as the sample becomes more $\alpha$-enhanced (Figure\,\ref{fig:afezscatter}). This result aligns with the ``upside-down" Galaxy formation model of \citet{bird13}, in which the thin and thick disk form during the smooth, continuous collapse of the star-forming gas reservoir, and the observations of a smoothly varying disk in \citet{bovy12b}.  
In addition, every star-forming epoch, which we associate with our bins in [$\alpha$/Fe], has consistent vertical metallicity structure. As noted in \citet{carraro98}, ``stars or clusters with different ages are not necessarily expected to trace the same metallicity gradient." The consistency over [$\alpha$/Fe] suggests similar star formation processes over the development of the disk. There is no clear discrepancy in vertical chemical structure between the proposed thin- and thick-disk components. 

Although our stellar sample supports a picture of a smoothly varying disk structure, it does not allow us to rule out either the multi- or single-component scenarios of Galactic disk development. However, our analysis of the vertical behavior of the Milky Way disk provides valuable constraints for Galaxy models experimenting with different evolutionary pathways, some with two distinct populations and others with a complex single population. We eagerly await upcoming data from Gaia-ESO \citep{gilmore12}, the Large Sky Area Multi-Object Fiber Spectroscopic Telescope (LAMOST; \citealt{cui12}), GALAH \citep{zucker12}, and APOGEE \citep{ahn13}, which may be better able to constrain the disk chemical and dynamical structure. 

\section{Summary}
\label{sec:summary} 

We have extracted a sample of $\sim$42,000 G-dwarf stars from the SEGUE survey and divided it into various chemical subsamples of approximately 1000 stars each. Thanks to the systematic target selection of SEGUE, we are able to account for the different biases in these subsamples, such that they are both unbiased and reflective of the properties of many more stars than the number observed spectroscopically (Tables\,\ref{tab:correlated_errors},\,\ref{tab:correlated_errors_overlap}). This allows us to determine an unbiased vertical metallicity gradient of a large number of stars over an unprecedented volume of the disk. In addition, the vertical metallicity gradients of the different chemical subsamples, associated with different epochs of star formation, indicate how the disk structure has changed over time. 

The vertical metallicity gradient is $-$0.243$^{+0.039}_{-0.053}$ dex kpc$^{-1}$ for the disk as a whole, which probes around 1400 stars with [Fe/H] between $-$1.000 and $-$0.035 and distances between 1.447 and 1.614 kpc (Table\,\ref{tab:correlated_errors}, Figure\,\ref{fig:vgrad_slopes}). This dataset covers the disk from heights above the Galactic plane from 0.3 to 1.6 kpc. While the measured vertical metallicity gradient is consistent with much of the literature, we expect that some of the discrepancies between different observational analyses are largely due to uncertainties in photometric metallicities, limited volume coverage, limited sample size, and inadequate correction for various target-selection biases. 

In contrast to the disk as a whole, we find small vertical metallicity gradients for individual [$\alpha$/Fe] populations ($<$ 0.1 dex in magnitude). There is negligible change in the median [Fe/H] with respect to increasing $|Z|$ for these subsamples; most are consistent with flat metallicity gradients within 2$\sigma$ (Figure\,\ref{fig:vgrad_slopes}). Similarly, we measure small vertical metallicity gradients when we divide the sample into larger bins of [$\alpha$/Fe], designed to exploit bimodality in [$\alpha$/Fe] vs. [Fe/H] and based on the picture of the chemically-separable thin and thick disk components (Figure\,\ref{fig:vgrad_slopes}). This result suggests there is little vertical change in median [Fe/H] throughout the disk at a given epoch of star formation. It also indicates that the vertical metallicity patterns of the disk are consistent over the full timeline of disk development, implying similar formation processes for both $\alpha$-rich and $\alpha$-poor disk stars. 

Comparison of the gradient of the total sample to those of individual [$\alpha$/Fe] populations over the same volume indicates that the varying $|Z|$ heights of each $\alpha$-population produces a strong gradient over the disk as a whole (Figures\,\ref{fig:afezscatter},\,\ref{fig:vgrad_slopes2}, and\,\ref{fig:mcpu3_hardcut_total}). More $\alpha$-poor stars are typically close to the plane of the Galaxy, with a limited range in $|Z|$. In contrast, $\alpha$-rich stars have a higher median $|Z|$ and appear to cover a wider range of Galactic heights (Figure\,\ref{fig:afezscatter}). Our measured radial metallicity gradient reflects the changing dominance of different $\alpha$-populations with height \citep{schlesinger12}, consistent with the various models that predict ``upside down" disk development (e.g., \citealt{bird13}, \citealt{stinson13}). 

Although the separability of the thin- and thick-disk component remains unclear, we find that the vertical metallicity gradient suggests consistent evolution processes for the different epochs of star formation throughout the disk. Similarly, we see continuous behavior of [Fe/H] with respect to $|Z|$ for the disk as a whole, which reflects the observed smooth variation in scale heights of different chemical populations. 

\acknowledgements 

We thank M. Hayden and J. Bovy for their helpful discussion on this work. 
We also thank the anonymous referee for their helpful comments. 
K.J.S. and J.A.J acknowledge partial support from NSF grant AST-0807997.  Y.S.L
and T.C.B. acknowledge partial support from grant PHY 08-22648:
Physics Frontiers Center/Joint Institute for Nuclear Astrophysics
(JINA), awarded by the U.S. National Science Foundation. J.C.B. acknowledges 
the support of the Vanderbilt Office of the Provost through the Vanderbilt Initiative 
in Data-intensive Astrophysics (VIDA).

Funding for SDSS-III has been provided by the Alfred P. Sloan
Foundation, the Participating Institutions, the National Science
Foundation, and the U.S. Department of Energy Office of Science. The
SDSS-III web site is http://www.sdss3.org/.

SDSS-III is managed by the Astrophysical Research Consortium for the
Participating Institutions of the SDSS-III Collaboration including the
University of Arizona, the Brazilian Participation Group, Brookhaven
National Laboratory, University of Cambridge, University of Florida,
the French Participation Group, the German Participation Group, the
Instituto de Astrofisica de Canarias, the Michigan State/Notre
Dame/JINA Participation Group, Johns Hopkins University, Lawrence
Berkeley National Laboratory, Max Planck Institute for Astrophysics,
New Mexico State University, New York University, Ohio State
University, Pennsylvania State University, University of Portsmouth,
Princeton University, the Spanish Participation Group, University of
Tokyo, University of Utah, Vanderbilt University, University of
Virginia, University of Washington, and Yale University.

\appendix

\section{Uncertainties in the Vertical Metallicity Gradient}
\label{appendix}

\subsection{Effect of the Radial Metallicity Gradient} 

There are two uncertainties that stem from our radial metallicity gradient correction. First, uncertainties in a star's chemistry and distance will result in a change in [Fe/H]. This effect is included as part of the uncertainties for each individual parameter, discussed in \S\,\ref{sec:unc_stellar_param}. Secondly, there are uncertainties on the radial metallicity gradients themselves, which will change the vertical gradient on a larger scale. We examine the vertical gradients for each subsample using the extreme positive and negative radial metallicity slope values reported by \citet{cheng12b}. 

\begin{figure}[htbp]
\begin{center}
\includegraphics[width=\textwidth]{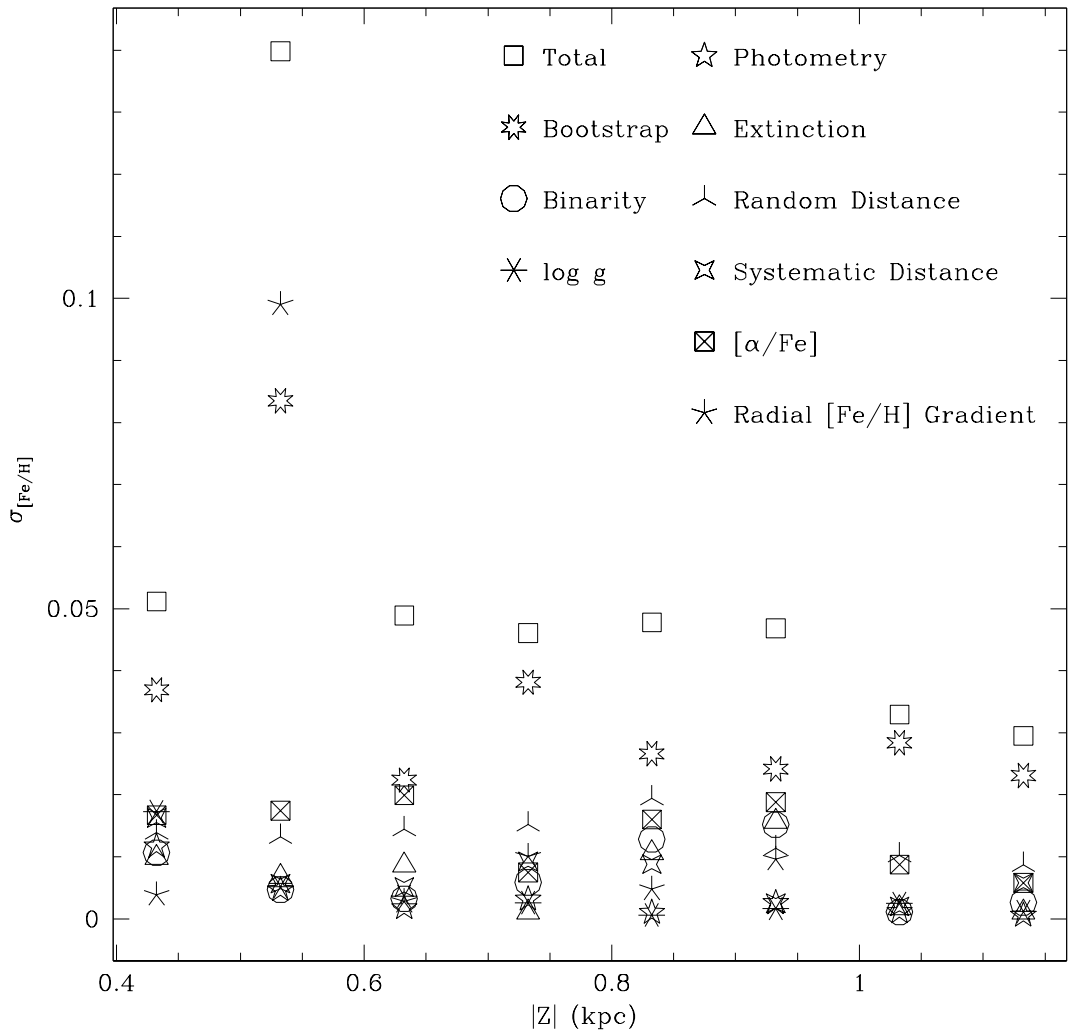}
\caption{The typical variation in [Fe/H] (in dex) at the range of $|Z|$ due to different uncertainties for our $\alpha$-bins 2 subsample, which has some of the largest uncertainties. For each subsample, there is some variation with respect to the relative importance of different sources of uncertainty. For example, the uncertainty of the radial metallicity gradient is large at low $|Z|$; thus, subsamples dominated by stars close to the Galactic plane have a larger contribution to uncertainty from this parameter than those at larger $|Z|$.  The largest uncertainty typically arises from our bootstrap errors, related to the weights and sample selection, for all of the subsamples. The uncertainties from undetected binarity, photometry, and extinction are generally negligible. 
}
\label{fig:vgrad_errors}
\end{center}
\end{figure}

At low $|Z|$, the uncertainties in the radial metallicity gradient are large, resulting in a larger uncertainty in the median [Fe/H] at this height ($\Delta$[Fe/H]$\approx$0.04 dex, see Figure\,\ref{fig:vgrad_errors}).  Above $|Z|$ = 0.8 kpc, variations in [Fe/H] from the radial metallicity gradient are negligible. The uncertainty in the radial metallicity gradient at low $|Z|$ also contributes uncertainty to our measured slopes. We determine the change in slope that occurs when we use the extreme values of the radial metallicity gradient, a mean change of 0.02 dex kpc$^{-1}$ over the different subsamples. 

\subsection{Uncertainties from Sample Selection}
\label{sec:bootstrap}

We employ a bootstrap analysis to examine the variation in the sample arising from uncertainties in atmospheric parameters, photometry, and target-selection weights. For our bootstrap, we randomly select G dwarfs, sampling with replacement, and compare the vertical metallicity gradients for 100 iterations of the sample. At a given $|Z|$, these uncertainties typically change the median [Fe/H] by  0.02 dex. At low $|Z|$, the stellar density is high and our sample size is small ($\approx$120 stars with $|Z|\leq$0.5 kpc), resulting in uncertainties of around 0.03 dex in [Fe/H] (see Figure\,\ref{fig:spatial_alpha}). Our $\alpha$-poor subsamples are generally at low $|Z|$; the bootstrapping errors for these samples are typically larger, with a maximum $\Delta$[Fe/H] of 0.1 dex. 

We quantify the change in slope due to uncertainties in sample selection by comparing the gradient for each bootstrapped iteration to our measured value, determining the 1$\sigma$ variation in slope. This approach typically causes the largest uncertainty in the gradient for each of the subsamples, as listed in Table\,\ref{tab:correlated_errors}.

\subsection{Uncertainties from Stellar Parameters}
\label{sec:unc_stellar_param} 

We use a Monte Carlo analysis of the Galaxy model of \citet{schonrich09a, schonrich09b} to examine the uncertainty in the vertical metallicity gradient due to errors in individual parameters (see \S\,\ref{sec:corr_errors}). For each source of SEGUE uncertainty, we convolve the expected uncertainty for each star (based on its $S/N$, [Fe/H], etc.) with a Gaussian 20 times to create 20 ``mock catalogs". We then apply the target-selection criteria, distance, and chemical cuts on the modelled lines of sight, modified to include uncertainties, to simulate each of our subsamples and examine how the resulting vertical metallicity gradient changes. For more details about modelling these uncertainties, see \citet{schlesinger12}. For each $|Z|$, we examine the [Fe/H] values, determining the 1$\sigma$ variation from the original model value. We also examine the variation in the vertical metallicity gradient slope over each iteration, determining the 1$\sigma$ uncertainty from the underlying model value. 

SEGUE uses color and magnitude cuts to identify the G-dwarf sample. Although the uncertainty in SDSS photometry is typically 2$-$3\%, stars could be erroneously included/excluded in the sample, as explored with our bootstrap analysis (\S\,\ref{sec:bootstrap}). These small photometric uncertainties have little effect on the vertical metallicity gradient; the uncertainty in slope due to photometry is around 0.005 dex kpc$^{-1}$. Undetected binarity could also shift the properties for a given star. However, \citet{schlesinger10} found that undetected binarity has little to no effect on [Fe/H] and photometry, implying that resulting uncertainties from binarity will be negligible. Using the methodology described in \citet{schlesinger12}, we simulate the change in stellar properties due to an undetected secondary. This has little effect on the vertical metallicity gradient, contributing an uncertainty of 0.006 dex kpc$^{-1}$ and a $\Delta$[Fe/H] at a given height of less than 0.01 dex. Finally, extinction can artificially shift stars in and out of the G-dwarf color and magnitude range. Values from \citet{sfd98} assume that each star lies behind the full amount of line-of-sight dust; this can lead to a reddening overcorrection, scattering blue stars in and red stars out of the appropriate color range, creating a metallicity bias. \citet{cheng12a} estimate the effect of differential reddening on SEGUE main-sequence turnoff stars over lines of sight at a range of extinctions. They demonstrated that in areas of low extinction, like the $< $0.2 mag in our different lines-of-sight, using \citet{sfd98} reddening estimates will not induce a bias in our sample against metal-rich stars by preferentially removing nearby stars. Following the methodology of \citet{schlesinger12}, we model the reddening with respect to distance for each line of sight using the data of \citet{cheng12a} in a Monte Carlo analysis and find that uncertainty from reddening leads to a mean error on the slope of around 0.007 dex kpc$^{-1}$. It is similarly small when examining the change in [Fe/H] at a given $|Z|$, less than 0.01 dex. 

Uncertainties associated with the estimates of atmospheric parameters are typically larger. The errors for the SSPP $\log g$ are $\pm$0.4 dex for $S/N$ of 25; for [$\alpha$/Fe] it is $\pm$0.10 at this $S/N$. These uncertainties can shift stars in and out of our G-dwarf sample, and our chemically-defined subsamples. Our Monte Carlo analysis of the \citet{schonrich09a, schonrich09b} model indicates that errors in $\log g$ and [$\alpha$/Fe] change the vertical metallicity gradients of our subsamples by less than 0.02 dex in [Fe/H] at a given $|Z|$ (see Figure\,\ref{fig:vgrad_errors}). These uncertainties also have a small affect on the measured slope (Table\,\ref{tab:correlated_errors}). 

\subsection{Distance Uncertainties} 

\citet{schlesinger12} performed an in depth analysis of the random and systematic distance errors using an isochrone-matching technique on this G-dwarf sample. The random distance uncertainty stems from errors in photometry, [Fe/H], [$\alpha$/Fe], and isochrone choice. There is also a systematic distance uncertainty from the age assumptions and undetected binarity. For metal-rich stars ([Fe/H]$ > -$0.5), the random uncertainty is $\approx$18$\%$ and the systematic uncertainty is $-$3$\%$. More metal-poor stars have a random distance uncertainty of $\approx$8$\%$ and a negligible systematic shift. 

Uncertainties in distance will shift stars in and out of our subsamples. In addition, they will affect a star's $|Z|$, and thus the vertical metallicity gradient. Using the same Monte Carlo technique explained above, the random distance uncertainties change the median [Fe/H] by approximately 0.02 dex at a given $|Z|$ and contribute around $\pm$0.01 dex kpc$^{-1}$ of uncertainty to our measured slopes. The systematic distance uncertainties have a smaller effect, $\pm$0.01 dex in [Fe/H] at a given $|Z|$ and a $\pm$0.007 dex kpc$^{-1}$ uncertainty in slope.

\end{document}